\documentclass[pageno]{jpaper} 

\usepackage{amsmath}

\usepackage{geometry}
\usepackage{times}
\usepackage{relsize}
\usepackage{enumerate}
\usepackage{enumitem}
\usepackage{graphicx}
\usepackage[usenames,dvipsnames]{xcolor}
\usepackage{xurl}

\hypersetup{colorlinks=true,
pdftitle={Pond},
pdfauthor={Submission 346},
pdfsubject={Memory Disaggregation},
pdfkeywords={CXL; Memory Disaggregation},
citecolor=Maroon,
linkcolor=Blue,
urlcolor=Maroon}

\usepackage{setspace}
\usepackage{rotating}
\usepackage{xspace}

\usepackage{floatflt}
\usepackage{wrapfig}
\usepackage{alltt}
\usepackage{epstopdf}
\usepackage{subcaption}  

\usepackage{comment}

\usepackage{fancyvrb}

\usepackage{etoolbox}
\apptocmd{\thebibliography}{\raggedright}{}{}

\usepackage[numbers,sort&compress]{natbib} 

\usepackage{colortbl} 

\usepackage{wasysym}  
\usepackage{pifont}   

\usepackage{upgreek}
\usepackage{amssymb} 

\usepackage{tikz, pgfplots, pgfplotstable}
\usetikzlibrary{patterns}
\usepgfplotslibrary{fillbetween}
\usepackage{upgreek}

\usepackage{threeparttop}

\setlist{noitemsep,leftmargin=*, topsep=4pt, partopsep=0pt}

\begin{document}


\newcommand{\myparagraph}[1]{\vfive\noindent\textbf{#1.}}

\newcommand*\circled[1]{\tikz[baseline=(char.base)]{
\node[shape=circle,fill,inner sep=.5pt] (char) {\textcolor{white}{\textbf{#1}}};}}

\def \rone {\circled{1}}
\def \rtwo {\circled{2}}

\def \us {$\upmu$s}  
\def \yes {$\surd$}  

\def \none {N$_1$}
\def \ntwo {N$_1$}
\def \ntri {N$_1$}
\def \nnnn {N$_n$}

\def \ra {$\rightarrow$}

\def \tms {$\times$}

\def \whitecircle {$\ocircle$}
\def \blackcircle {\ding{108}}
\def \whitesquare {$\Box$}
\def \blacksquare {\ding{110}}
\def \whitediamond {$\Diamond$}
\def \blackdiamond {\ding{117}}
\def \whitetriangle {$\bigtriangleup$}
\def \blacktriangle {\ding{115}}
\def \whitedtriangle {$\bigtriangledown$}
\def \blackdtriangle {\ding{116}}

\newcommand{\tc}[1]{\textbf{\textcircled{#1}}}
\def \tcone {\tc{1}\xspace}
\def \tctwo {\tc{2}\xspace}
\def \tctri {\tc{3}\xspace}
\def \tcfour {\tc{4}\xspace}
\def \tcfive {\tc{5}\xspace}

\def \cm {\checkmark}
\def \xm {\ding{55}}


\newcommand{\vtwenty}{\vspace{20pt}}
\newcommand{\vfifteen}{\vspace{15pt}}
\newcommand{\vten}{\vspace{10pt}}
\newcommand{\vfive}{\vspace{5pt}}
\newcommand{\vtri}{\vspace{3pt}}
\newcommand{\vtwo}{\vspace{2pt}}

\newcommand{\vminfive}{\vspace{-5pt}}
\newcommand{\vminten}{\vspace{-10pt}}
\newcommand{\vminfifteen}{\vspace{-15pt}}
\newcommand{\vmintwenty}{\vspace{-20pt}}

\def \hmina {\hspace{-0.1in}}
\def \hminb {\hspace{-0.2in}}

\newcommand{\ub}[1]{\underline{{\bf #1}}}
\newcommand{\ts}[1]{{\tt{\small#1}}}
\newcommand{\bquote}{\vspace{-0.25cm} \begin{quote}}
\newcommand{\equote}{\end{quote}\vspace{-0.2cm} }
\def \sec {\S}
\def \yes {$\surd$}

\def \nospace {
  \setlength{\itemsep}{0pt}
  \setlength{\parskip}{0pt}
  \setlength{\parsep}{0pt}
}

\newcommand{\border}[1]{\textbf{\textcolor{Magenta}{[BDR]#1}}}
\renewcommand{\border}[1]{}

\newcommand{\myquote}[1]{
\begin{quote}
\centering
\small
\textit{#1}
\end{quote}
}

\newenvironment{enumerate2}{
  \begin{enumerate} \vminfive
  \setlength{\itemsep}{1pt}
  \setlength{\parskip}{0pt}
  \setlength{\parsep}{0pt}
}{
  \end{enumerate}
}

\newenvironment{itemize2}{
  \begin{itemize} \vminfive
 \renewcommand{\labelitemi}{-}
  \setlength{\itemsep}{1pt}
  \setlength{\parskip}{0pt}
  \setlength{\parsep}{0pt}
}{
  \end{itemize}
}


\newcommand{\note}[1]{\textcolor{darkgray}{{\footnotesize {\em (Note: #1)}}}}

\newcommand{\student}[1]{\textcolor{purple}{{\footnotesize {\bf (STU: #1)}}}}

\newcounter{hsgcounter}
\newcommand{\hsg}[1]{{\footnotesize
\textbf{\textcolor{red}{(HSG$_{\arabic{hsgcounter}}$: #1)}}}
\stepcounter{hsgcounter}}

\newcommand{\revs}[1]{\textcolor{blue}{\textit{(REVS: #1)}}}
\newcommand{\hcl}[1]{{\footnotesize\textbf{\textcolor{magenta}{        [ HCL: #1 ]}}}}
\newcommand{\dsb}[1]{{\footnotesize\textbf{\textcolor{CornflowerBlue}{          [ DSB: #1 ]}}}}
\newcommand{\anb}[1]{{\footnotesize\textbf{\textcolor{CornflowerBlue}{ [ ANB: #1 ] }}}}
\newcommand{\rb}[1]{{\footnotesize\textbf{\textcolor{Magenta}{ [ RB: #1 ] }}}}
\newcommand{\stn}[1]{{\footnotesize\textbf{\textcolor{blue}{           [ STN: #1 ] }}}}
\newcommand{\dkp}[1]{{\footnotesize\textbf{\textcolor{red}{           [ DKP: #1 ] }}}}
\newcommand{\irz}[1]{{\footnotesize\textbf{\textcolor{red}{           [ IRZ: #1 ] }}}}

\newcommand{\pc}[1]{\textcolor{blue}{\textit{(PC: #1)}}} 
\newcommand{\todo}[1]{\textcolor{red}{{\footnotesize {\bf (TODO: #1)}}}}

\definecolor{myblue}{rgb}{0.32, 0.5, 0.84}
\newcommand{\newtxt}[1]{#1} 
\newcommand{\oldtxt}[1]{\textcolor{gray}{{\footnotesize {\em OLD TEXT: #1}}}}
\newcommand{\bluetxt}[1]{\textcolor{blue}{#1}}
\newcommand{\rbt}[1]{\textcolor{red}{\textbf{#1}}}
\newcommand{\bbt}[1]{\textcolor{blue}{\textbf{#1}}}




\def \vvvnb {\vfifteen \noindent $\bullet$~}
\def \vvnb {\vten \noindent $\bullet$~}
\def \vnb {\vfive \noindent $\bullet$~}
\def \vn {\vfive \noindent}

\def \mb {\vspace{8pt}\nb}
\def \tb {\vspace{8pt}\nb}

\def \vvni {\vten \noindent}
\def \vni {\vfive \noindent}
\def \nb {\noindent $\bullet$~}
\def \ni {\noindent}
\def \bb {$\bullet$~}

\newcommand{\hypo}[1]{
\begin{quote}
\stepcounter{HYPO}{\bf Hypothesis \arabic{HYPO}:}
{\em #1}
\end{quote}
}

\newcommand{\taskformat}[2]{#1\textsc{#2}}

\newcommand{\task}[3]{
\begin{quote}
\phantomsection
\hypertarget{task#1#2}{}
{\bf Task \taskformat{#1}{#2}:}
{\em #3}
\end{quote}
}

\newcommand{\tasklink}[2]{\hyperlink{task#1#2}{\taskformat{#1}{#2}}}

\newcounter{HYPO}
\newcounter{TASK}

\newcommand{\rs}{{ResearchStaff$_1$}}
\newcommand{\pd}{{\bf Postdoc$_1$}}
\newcommand{\raOne}{{\bf RA$_1$}}
\newcommand{\raTwo}{{\bf RA$_2$}}
\newcommand{\ndv}{{\bf NDV}}
\newcommand{\ug}{{\bf Undergrad$_1$}}


\newcommand{\sssubsection}[1]{\vten\ni\textbf{\large{\textsc{#1}}}}

\newcounter{mysubcounter}
\setcounter{mysubcounter}{1}
\newcommand{\mysub}[1]{\vten\noindent\textcolor{black}{\textbf{\textbf{#1}\stepcounter{mysubcounter}}}}

\newcommand{\emptypage}{
\newpage
(empty page)
}

\newcommand{\myrotate}[1]{\begin{rotate}{90} {\bf #1} \end{rotate}}

\newcommand{\mycaption}[3]{
\begin{spacing}{0.95}
\caption{
\label{#1}
{\small \bf #2. }
{\it \small #3}
}
\end{spacing}
}

\newcommand{\mysubcaption}[3]{
\begin{spacing}{0.95}
\vminfifteen
\caption{
\label{#1}
{\bf #2} 
{\em \small #3} 
}
\end{spacing}
}

\newcommand{\eg}{\textit{e.g.}\xspace}
\newcommand{\ie}{\textit{i.e.}\xspace}
\newcommand{\etal}{\textit{et al.}\xspace}
\newcommand{\etc}{etc.}
\def \th {$^{th}$}

\newcommand{\sstar}{$^{*}$}
\newcommand{\stwostars}{$^{**}$}
\newcommand{\stristars}{$^{***}$}
\newcommand{\srealstar}{$^{\star}$}
\newcommand{\sdag}{$^{\dag}$}
\newcommand{\sddag}{$^{\ddag}$}
\newcommand{\splus}{$^{+}$}
\newcommand{\scircle}{$^{\circ}$}

\newcounter{Xcounter}
\newcommand{\xxxreset}{\setcounter{Xcounter}{1}}
\newcommand{\xxx}{{\footnotesize
\textcolor{red}{
\textbf{xxx$_{\arabic{Xcounter}}$}\stepcounter{Xcounter}}~}}

\newcommand{\xxxinfig}{\textcolor{red}{\textbf{xx}}} 


\newcounter{Fcounter}
\newcommand{\freset}{\setcounter{Fcounter}{1}}

\newcommand{\finding}[1]{
\begin{spacing}{0.80}
\findingTable{#1}
\end{spacing}
}

\definecolor{fgray}{gray}{0.9}

\newcommand{\findingTable}[1]{
\begin{table}[h!]
\begin{tabular}{|p{3.2in}|}
\hline
\rowcolor{fgray}
\findingBody{#1}\\
\hline
\end{tabular}
\end{table}
\vminten
}

\newcommand{\findingBody}[1]{{\small
\textbf{Finding \#$\arabic{Fcounter}$:}
\stepcounter{Fcounter}
#1 }
}

\newcommand{\myfinding}[1]{\vspace*{5pt}\noindent{{\bf{Finding \arabic{Fcounter}.}\stepcounter{Fcounter}}}}

\setcounter{Fcounter}{1}

\newcommand{\supsection}[1]{\noindent{\Large{\bf #1}}\vten}

\newcommand{\summaryheader}{
  \thispagestyle{empty}
  \noindent {\bf\Large B $\;$ Project Summary}
}

\newcommand{\bodyheader}{
  \setcounter{page}{1}
  \noindent {\bf\Large D $\;$ Project Description}
}




\def \numTotalApps {158}
\def \maxSlowdown {51\%}

\def \ione {I$_1$}

\newcommand{\sys}{Pond\xspace}

\newcommand{\azure}{Azure\xspace}

\newcommand{\cvn}{cvNUMA\xspace}
\renewcommand{\cvn}{zNUMA\xspace}

\newcommand{\pdm}{\texttt{PDM}\xspace}
\newcommand{\tp}{\texttt{TP}\xspace}

\def \maxNodes {60}

\def \numItems {\xxx\xspace}


\def \mytitle {\mbox{\sys: CXL-Based Memory Pooling Systems for Cloud Platforms}}

\title{\bf \mytitle}

%


\def \ispace {\textcolor{white}{\rule{1.8in}{1pt}}}

\author{Huaicheng Li\sdag, Daniel S. Berger\sstar\sddag, Stanko Novakovic\sstar, Lisa Hsu\sstar, Dan Ernst\sstar, \\ Pantea Zardoshti\sstar, Monish Shah\sstar, Samir Rajadnya\sstar, Scott Lee\sstar, Ishwar Agarwal\sstar,\\
Mark D. Hill\sstar\scircle, Marcus Fontoura\sstar, Ricardo Bianchini\sstar \\ \sdag {\small Virginia Tech and CMU}\quad \sstar {\small Microsoft Azure}\quad \sddag {\small University of Washington}\quad \scircle {\small University of Wisconsin-Madison}}

\date{}
\maketitle




\def \myabstract{%
Using data from a public cloud provider, we find that up to 25\%
of memory is \emph{stranded}, \ie, it is leftover after the servers'
cores have been rented to VMs.  Memory disaggregation promises
to reduce this stranding.  However, making
disaggregation practical for production cloud deployment remains
challenging.  For example, RDMA-based disaggregation involves too much
overhead for latency-sensitive workloads and transparent latency
management is incompatible with
virtualization acceleration.  The emerging Compute Express Link (CXL)
standard offers a low-overhead substrate upon which we can build
memory disaggregation while overcoming these challenges.  Thus, in
this paper, we propose \sys, a full-stack CXL-based disaggregation
system that meets the requirements of cloud providers.  \sys includes
a memory pool controller, and prediction-based system software and
distributed control plane designs.  Its predictions of VM latency
sensitivity and memory usage allow it to nearly perfectly split
workloads across local and pooled memory, mitigating the higher pool
latency.

Our analysis of production clusters shows that small pools of 8-16
sockets are sufficient to reduce stranding significantly.  It also
shows that $\sim$50\% of all VMs never touch 50\% of their rented
memory.  From emulated experiments with 150+ workloads and \sys's
memory split, we show that pooling incurs a configurable performance
loss between 1-5\%.  Finally, we show that \sys can achieve a 9-10\% reduction in
overall DRAM, which represents hundreds of millions of dollars
in cost savings for a large cloud provider.
}

\def \myabstract{%
Cloud providers seek to achieve stringent performance requirements and low cost.
DRAM has emerged as a driver of cost and memory pooling promises to improve DRAM utilization.
However, pooling is challenging under cloud performance requirements.
This paper proposes \sys, the first memory pooling system that achieves cloud performance goals and effectively reduces DRAM cost.
\sys builds on the emerging Compute Express Link (CXL) interface and two insights.
First, our analysis shows that grouping 4-16 dual-socket (or 8-32 single-socket) servers is enough to achieve most of the benefits of pooling.
This enables a small-pool design with low access latency.
Second, \sys introduces novel machine learning models that can effectively predict how much NUMA-local and pool memory to allocate to a VM to resemble NUMA-local memory performance.
%
Our evaluation with 150+ workloads shows that \sys only incurs 1-5\%
configurable performance loss while improving DRAM efficiency by
9-10\%.
%
}

\def \myabstract{%
Public cloud providers seek to meet stringent performance requirements and low hardware cost.
A key driver of performance and cost is main memory.
Memory pooling promises to improve DRAM utilization and thereby reduce costs.
However, pooling is challenging under cloud performance requirements.
This paper proposes \sys, the first memory pooling system that both meets cloud performance goals and significantly reduces DRAM cost.
\sys builds on the Compute Express Link (CXL) standard for load/store access to pool memory and two key insights.
First, our analysis of cloud production traces shows that pooling across 8-16 sockets is enough to achieve most of the benefits.
This enables a small-pool design with low access latency.
Second, it is possible to create machine learning models that can accurately predict how much local and pool memory to allocate to a virtual machine (VM) to resemble same-NUMA-node memory performance.
%
Our evaluation with 158 workloads shows that \sys reduces DRAM costs by
7\% with performance within 1-5\% of same-NUMA-node VM allocations.
%
}

\begin{abstract}
\ni\textit{\myabstract}
\end{abstract}

\section{Introduction}
\label{sec-new-intro}

\myparagraph{Motivation}
Many public cloud customers deploy their workloads in the form of
virtual machines (VMs), for which they get virtualized compute with
performance approaching that of a dedicated cloud, but without having
to manage their own on-premises datacenter. This creates a major
challenge for public cloud providers: achieving excellent performance
for opaque VMs (\ie, providers do not know and should not inspect what
is running inside the VMs) at a competitive hardware cost.

A key driver of both performance and cost is main memory.  The gold
standard for memory performance is for accesses to be served by the
same NUMA node as the cores that issue them, leading to latencies in
tens of nanoseconds.  A common approach is to
preallocate all VM memory on the same NUMA node as the VM's cores.
Preallocating and statically pinning memory also facilitate the use
of virtualization
accelerators~\cite{nicpagefault.asplos17,tian2020coiommu,yassour2010dma,willmann2008protection,amit2011viommu,ben2010turtles},
which are enabled by default, for example, on AWS and
Azure~\cite{awsaccelnet,azureaccelnet}. At the same time,
DRAM has become a major portion of hardware cost due to its poor
scaling properties with only nascent
alternatives~\cite{memscaling.imw13, archshield.isca13, mutlu2015main,
  dramscaling.imw15, dramscalingchallenges.imw20,
  nextnewmemories.web19,micron3dxp.news21}.  For example, DRAM can be
50\% of server cost~\cite{cxlandgenz.web20}.

Through analysis of production traces from \azure,
we identify {\em memory stranding} as a dominant source of memory waste and a potential source of massive cost savings.
Stranding happens when all cores of a server are rented (\ie, allocated to customer VMs) but unallocated memory capacity remains and cannot be rented.
We find that up to 25\% of DRAM becomes stranded as more cores become allocated to VMs.

\myparagraph{Limitations of the state of the art} Despite this
significant amount of stranding, reducing DRAM usage in the public
cloud is challenging due to its stringent performance requirements.
For example, existing techniques for process-level memory
compression~\cite{softfarmem.asplos19,weiner2022tmo} require page
fault handling, which adds microseconds of latency, and moving away
from statically preallocated memory.

Pooling memory via memory disaggregation is a promising approach
because stranded memory can be returned to the disaggregated pool and
used by other servers.  Unfortunately, existing pooling systems also
have microsecond access latencies and require page
faults~\cite{memblade.isca09, lim2011disaggregated,
  orchdisaggmem.tc20, sysdisaggmem.hpca12, nicpagefault.asplos17,
  aifm.osdi20, angel2020disaggregation} or changes to the VM
guest~\cite{aifm.osdi20, semeru.osdi20, kona.asplos21, farm.nsdi14,
  remoteregions.atc18, kvdisagg.atc20, farview.corr21,
  sysdisaggmem.hpca12, resdisagg.osdi16, infiniswap.nsdi17,
  legoos.osdi18, softfarmem.asplos19, dcm.tc19, fastswap.eurosys20,
  leap.atc20, fluidmem.icdcs20, orchdisaggmem.tc20,ememdisagg.corr20}.

\myparagraph{Our work} This work describes \sys, {\em the first system
  to achieve both same-NUMA-node memory performance and competitive
  cost for public cloud platforms.}  To achieve this, \sys combines
hardware and systems techniques.  It relies on the Compute Express
Link (CXL) interconnect standard~\cite{cxlsite.web20}, which enables
cacheable load/store (\texttt{ld/st}) accesses to pooled memory on
Intel, AMD, and ARM processors~\cite{intelsapphire.web21, arm2021cxl,
  amdgenoa} at nanosecond-scale latencies.  CXL access via
loads/stores is a game changer as it allows memory to remain
statically preallocated while physically being located in a shared
pool.  However, even with loads/stores, CXL accesses still face higher
latencies than same-NUMA-node accesses.  \sys introduces systems
support for CXL-based pooling that dramatically
reduces the impact of this higher latency.

\sys is feasible because of four key insights.
First, by analyzing traces from 100 production clusters at \azure, we find that pool sizes between 8-16 sockets lead to sufficient DRAM savings.
The pool size defines the number of CPU sockets able to use pool memory.
Further, analysis of CXL topologies lead us to estimate that CXL will add 70-90ns to access latencies over same-NUMA-node DRAM with a pool size of 8-16 sockets, and add more than 180ns for rack-scale pooling.
We conclude that grouping 8 dual-socket (or 16 single-socket) servers is enough to achieve most of the benefits of pooling.

\newtxt{Second, by emulating either 64ns or 140ns of memory access overheads, we find that 43\% and 37\% of 158 workloads are within 5\% of the performance on same-NUMA-node DRAM when entirely allocated in pool memory.
However, more than 21\% of workloads suffer a performance loss above 25\%.}
This emphasizes the need for small pools and shows the challenge with achieving same-NUMA-node performance.
This characterization also allows us to train a machine learning (ML) model that can identify a subset of insensitive workloads ahead of time to be allocated on the \sys memory pool.

Third, we observe through measurements at \azure that $\sim$50\% of all VMs touch less than 50\% of their rented memory.
Conceptually, allocating untouched memory from the pool should not have any performance impact even for latency-sensitive VMs.
We find that --- while this concept does not hold for the uniform address spaces assumed in prior work~\cite{memblade.isca09, lim2011disaggregated, orchdisaggmem.tc20, sysdisaggmem.hpca12, nicpagefault.asplos17, aifm.osdi20, angel2020disaggregation} --- it does hold if we expose pool memory to a VM's guest OS as a {\underline{z}ero-core virtual NUMA ({\bf \cvn}) node}, \ie, a node with memory but no cores, like Linux's CPU-less NUMA~\cite{cpulessnuma.web19}.
Our experiments show \cvn effectively biases memory allocations away from the \cvn node.
Thus, a VM with a \cvn sized to match its untouched memory will indeed not see any performance impact.

Fourth, \sys can allocate CXL memory with same-NUMA-node performance using correct predictions of {\bf a)} whether a VM will be latency-sensitive and {\bf b)} a VM's amount of untouched memory.
For incorrect predictions, \sys introduces a novel monitoring system that detects poor memory performance and triggers a mitigation that migrates the VM to use only same-NUMA-node memory.
Further, we find that all inputs to train and run \sys's ML models can be obtained from existing hardware telemetry with no measurable overhead.

%


\myparagraph{Artifacts}
CXL is still a year from broad deployment.
Meanwhile, deploying \sys requires extensive testing within \azure's system software and distributed software stack.
We implement \sys on top of an emulation layer that is deployed on production servers.
This allows us to prove the key concepts behind \sys by exercising the VM allocation workflow, \cvn, and by measuring guest performance.
Additionally, we support the four insights from above by reporting from extensive experiments and measurements in \azure's datacenters. We evaluate the effectiveness of pooling using simulations based on VM traces from 100 production clusters.

\myparagraph{Contributions} Our main contributions are:
\begin{itemize}
\item The first public characterization of memory stranding and untouched memory at a large public cloud provider.
\item The first analysis of the effectiveness and latency of different CXL memory pool sizes.
\item \sys, the first CXL-based full-stack memory pool that is practical and performant for cloud deployment.
\item An accurate prediction model for latency and resource management at datacenter scale. These models enable a configurable performance slowdown of 1-5\%.
\item \newtxt{An extensive evaluation that validates \sys's design including the performance of \cvn and our prediction models in a production setting.
  Our analysis shows that we can reduce DRAM needs by 7\% with a \sys pool spanning 16 sockets, which corresponds to hundreds of millions of dollars for a large cloud provider.}
\end{itemize}

\begin{figure}[t!]
  \centering
\includegraphics[width=\columnwidth]{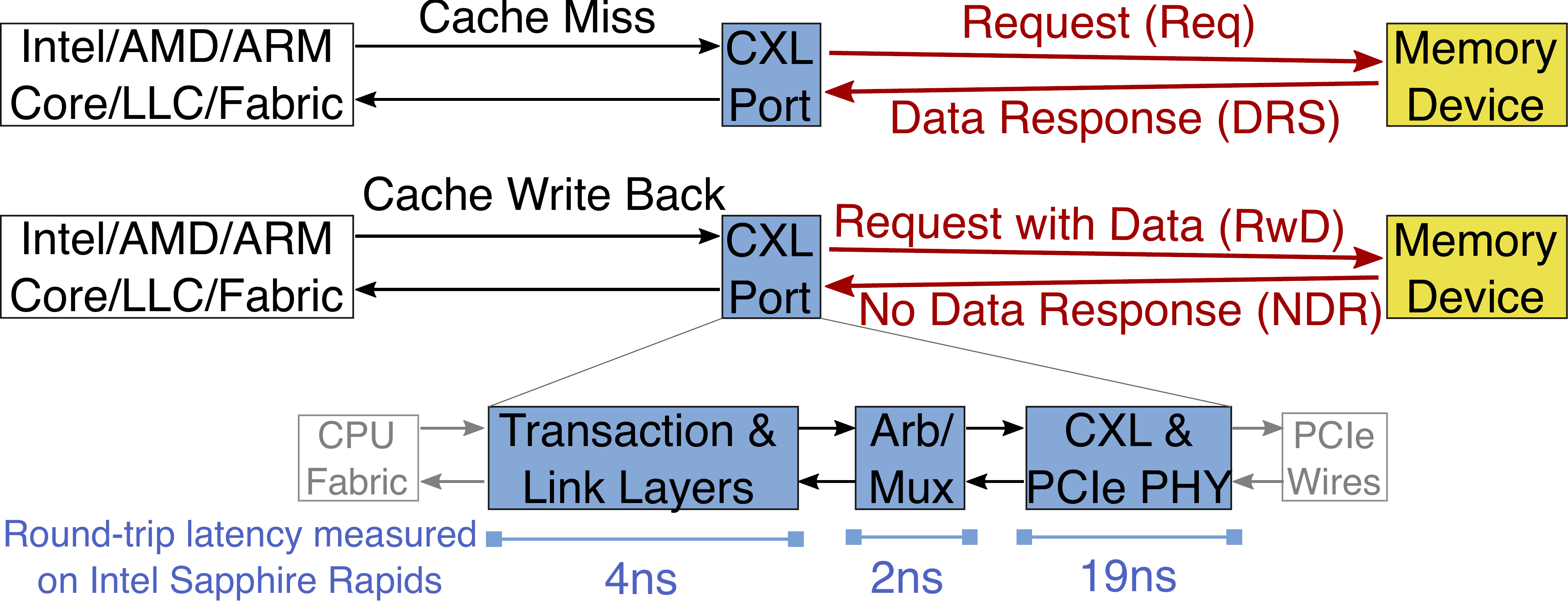}
\vminten

\mycaption{fig-cxlreadwrite}{CXL Request Flow
    (\sec\ref{sec-bg})}{\newtxt{CPU cache misses and write-backs to
    addresses mapped to CXL devices are translated to requests on a CXL
    port by the HDM decoder. Intel measures the round-trip port latency
    to be 25ns.}}

\end{figure}

\section{Background}
\label{sec-bg}

{\bf Hypervisor memory management.}
Public cloud workloads are virtualized~\cite{firecracker.nsdi20}.
To maximize performance and minimize overheads, hypervisors perform minimal memory management and rely on
virtualization accelerators to improve I/O performance~\cite{intelvtd.web20,nicpagefault.asplos17,leapio.asplos20,sriov}.
Examples of common accelerators are direct I/O device assignment
(DDA)~\cite{intelvtd.web20,nicpagefault.asplos17} and Single Root I/O
Virtualization (SR-IOV)~\cite{leapio.asplos20,sriov}.
Accelerated networking is enabled by default on AWS and Azure~\cite{awsaccelnet,azureaccelnet}.
As pointed out in prior work, virtualization acceleration requires statically preallocating (or ``pinning'') a VM's entire address space~\cite{nicpagefault.asplos17,tian2020coiommu,yassour2010dma,willmann2008protection,amit2011viommu,ben2010turtles}.

\myparagraph{Memory stranding} Cloud VMs demand a vector of resources
(\eg, CPUs, memory, \etc)
~\cite{resourcecentral.sosp17, hadary2020protean,
googlejobpacking.cluster14, borg.eurosys15}.
Scheduling VMs thus leads to a multi-dimensional bin-packing
problem~\cite{binpackheuristics.web11, tetris.sigcomm14, hadary2020protean, bpbounds.stoc13}
which is complicated by constraints such as spreading VMs across multiple failure
domains.
Consequently, it is difficult to provision servers that closely
match the resource demands of the incoming VM mix.
When the DRAM-to-core ratio of VM arrivals and the server resources
do not match, tight packing becomes more difficult.
We define a resource as \emph{stranded} when
it is technically available to be rented to a customer, but is
practically unavailable as some other resource has exhausted. The typical scenario for {\em
memory stranding} is that all cores have been
rented, but there is still memory available in the server.

\myparagraph{Reducing stranding} Multiple
techniques can reduce memory stranding. For example,
oversubscribing cores~\cite{smartharvest.eurosys21,harvestslo.osdi20} enables more memory to be rented.
However, oversubscription only applies to a subset of VMs for performance reasons.
Our measurements at \azure (\sec\ref{sec-strand}) include
clusters that enable oversubscription and still show significant memory stranding.

The approach we target is to disaggregate a
portion of memory into a pool that is accessible by multiple
hosts~\cite{resdisagg.osdi16, fastnetdisagg.socc17,
thymesisflow.micro20}. This breaks the fixed hardware
configuration of servers.
By dynamically reassigning memory to different
hosts at different times, we can shift memory resources to where they
are needed, instead of relying on each individual server to be configured
for all cases pessimistically.
Thus, we can provision servers close to the average
DRAM-to-core ratios and tackle deviations via the memory pool.

\begin{figure}[t!]
\begin{center}
\includegraphics[width=.9\columnwidth]{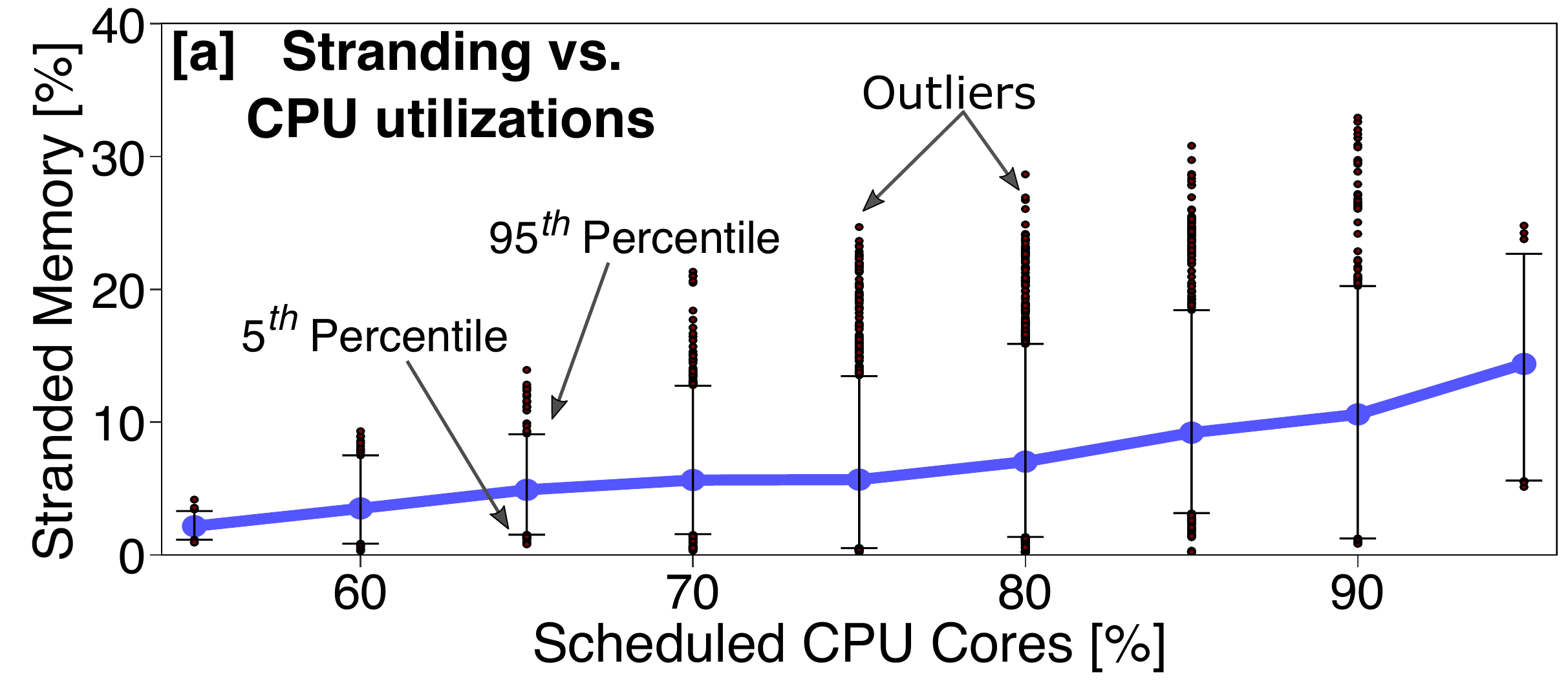}
{\hspace*{5pt}\includegraphics[width=.93\columnwidth]{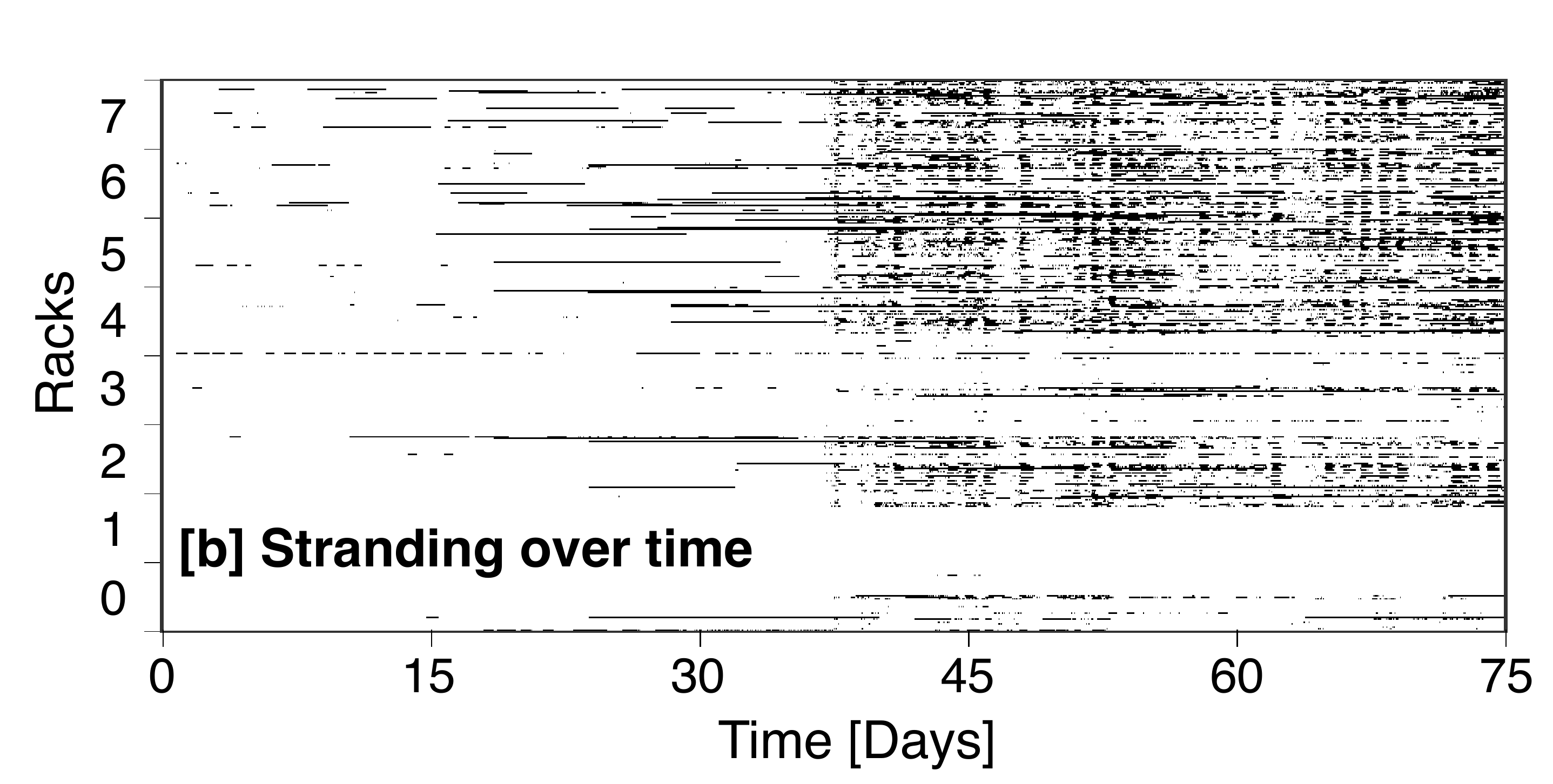}}
    \mycaption{fig-stranding-bars}{Memory stranding}{(a) Stranding increases significantly as
    more CPU cores are scheduled; (b) Stranding changes dynamically over time.}
\vminfive
\end{center}
\end{figure}

\myparagraph{Pooling via CXL}
\newtxt{CXL contains multiple protocols including \texttt{ld/st} memory semantics (CXL.mem) and I/O semantics (CXL.io).
CXL.mem maps device memory to the system address space.
Last-level cache (LLC) misses to CXL memory addresses translate into requests on a CXL port whose reponses bring the missing cachelines (Figure~\ref{fig-cxlreadwrite}).
Similarly, LLC write-backs translate into CXL data writes.
Neither action involves page faults or DMAs.
CXL memory is virtualized using hypervisor page tables and the memory-management unit and is thus compatible with virtualization acceleration.
The CXL.io protocol facilitates device discovery and configuration.
CXL 1.1 targets directly-attached devices, 2.0~\cite{cxl2spec.web20,cxl2whitepaper.web21} adds switch-based pooling, and 3.0~\cite{debendra2022fms,cxl3spec} standardizes switch-less pooling (\S\ref{sec-des}) and higher bandwidth.}

\newtxt{CXL.mem uses PCIe's eletrical interface with custom link and transaction layers for low latency.
With PCIe 5.0, the bandwidth of a birectional \tms{8}-CXL port at a typical 2:1 read:write-ratio matches a DDR5-4800 channel.
CXL request latencies are largely determined by the CXL port.
Intel measures round-trip CXL port traversals at 25ns~\cite{debendra2022hoti} which, when combined with expected controller-side latencies, leads to an end-to-end overhead of 70ns for CXL reads, compared to NUMA-local DRAM reads.
While FPGA-based prototypes report higher latency~\cite{maruf2022tpp,gouk2022direct}, Intel's measurements match industry-expectations for ASIC-based memory controllers~\cite{maruf2022tpp,debendra2022hoti,cxl3spec}.}

\section{Memory Stranding \& Workload Sensitivity to Memory Latency}
\label{sec:measurements}

\subsection{Stranding at \azure}
\label{sec-strand}

This section quantifies the severity of memory stranding
and untouched memory at \azure using production data.

\myparagraph{Dataset} We measure stranding in 100 cloud clusters over a
75-day period.  These clusters host mainstream first-party and third-party VM workloads.
They are representative of the majority of the server fleet.
We select clusters with similar deployment years, but spanning all major
regions on the planet. A trace from each cluster contains millions of per-VM
arrival/departure events, with the time, duration, resource demands, and
server-id.

\begin{figure}[t!]
\begin{center}
\includegraphics[width=.85\columnwidth]{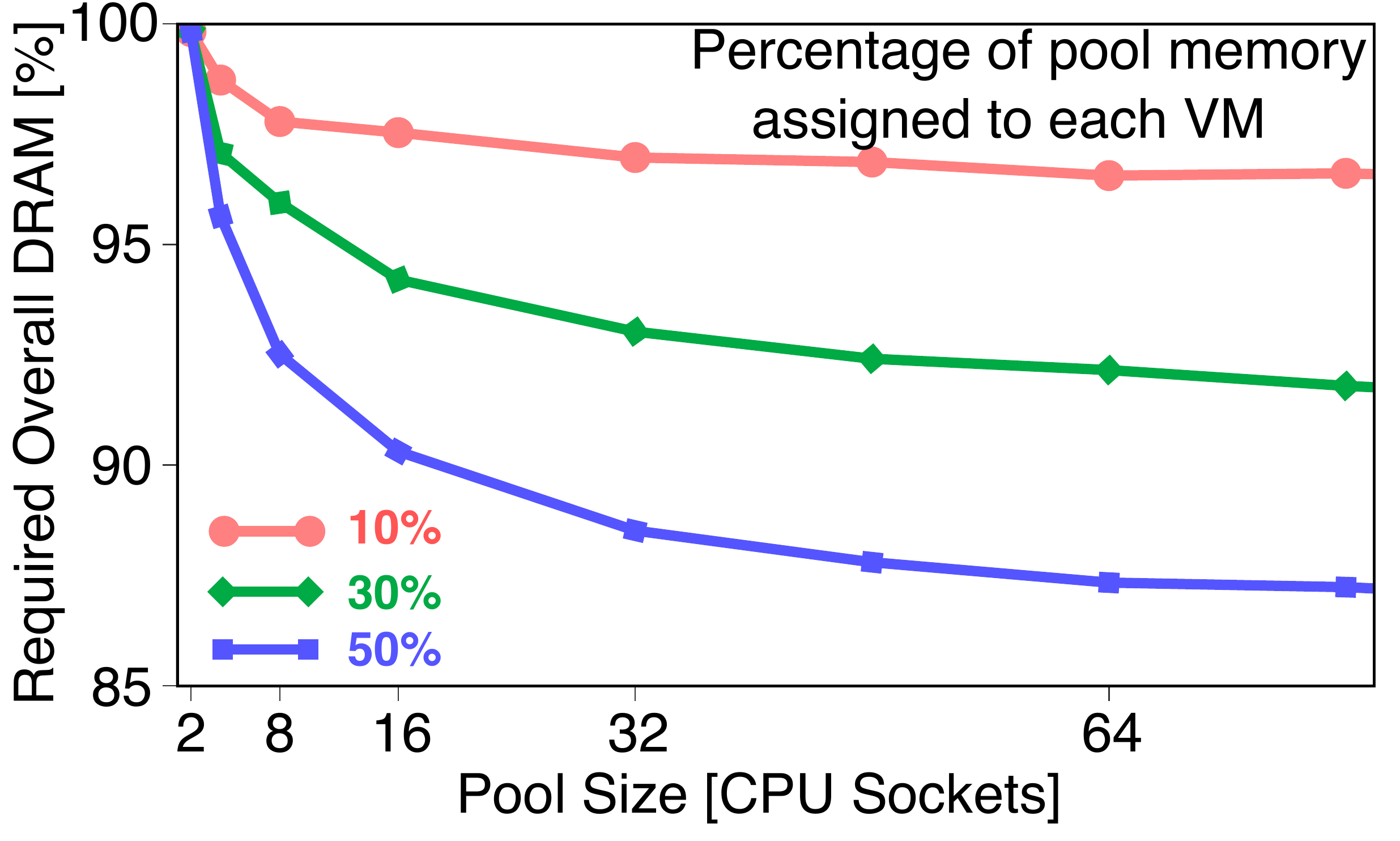}
\mycaption{fig-stranding-pooling}{Impact of pool size (\sec\ref{sec-strand})}{Small pools of
32 sockets are sufficient to significantly reduce memory needs.}
\vminfive
\end{center}
\end{figure}

\def \hmina {\hspace{0.02in}}
\def \hminb {\hspace{-0.2in}}

\def \fgw {6.5in}
\def \fgh {1.085in}


\begin{figure*}[ht!]
\centerline{
\includegraphics[width=\fgw]{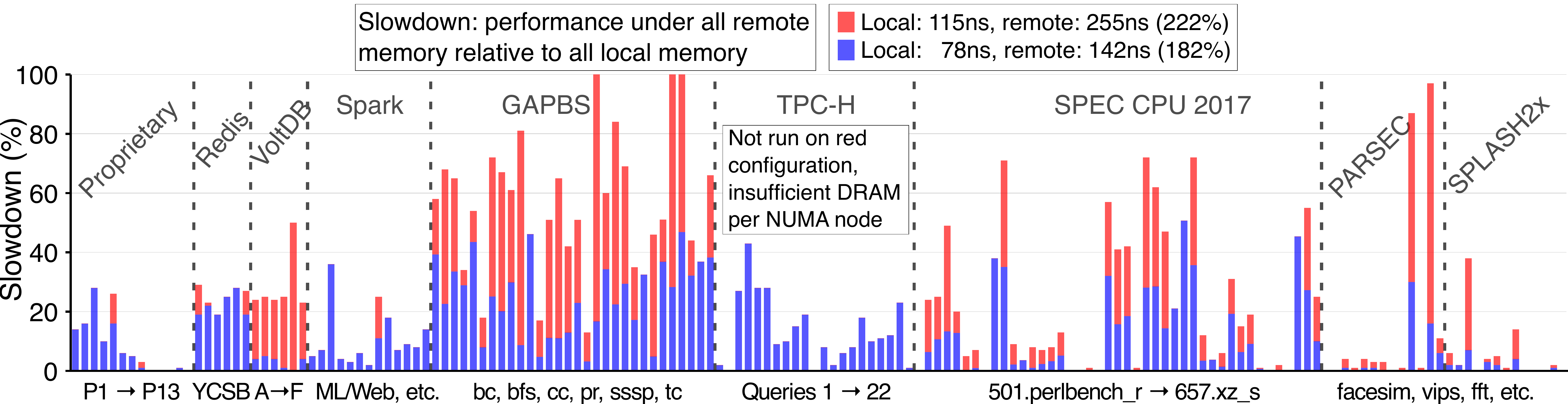}
}

%
\mycaption{fig-cxl}{Performance slowdowns when memory latency increases by 182-222\% (\sec\ref{sec-mot-cxl})}{\newtxt{Workloads have different sensitivity to additional memory latency (as in CXL).
X-axis shows \numTotalApps\ representative workloads;
Y is the normalized performance slowdown, \ie, performance under higher (remote) latency relative to all local memory.
``Proprietary'' denotes production workloads at \azure.\vten
%
}}
%

\end{figure*}

\def \hmina {\hspace{-0.4in}}
\def \hminb {\hspace{-0.35in}}

\def \fgw {3.2in}
\def \fgh {1.085in}

\begin{figure}[t!]
\centerline{
\includegraphics[width=.88\columnwidth]{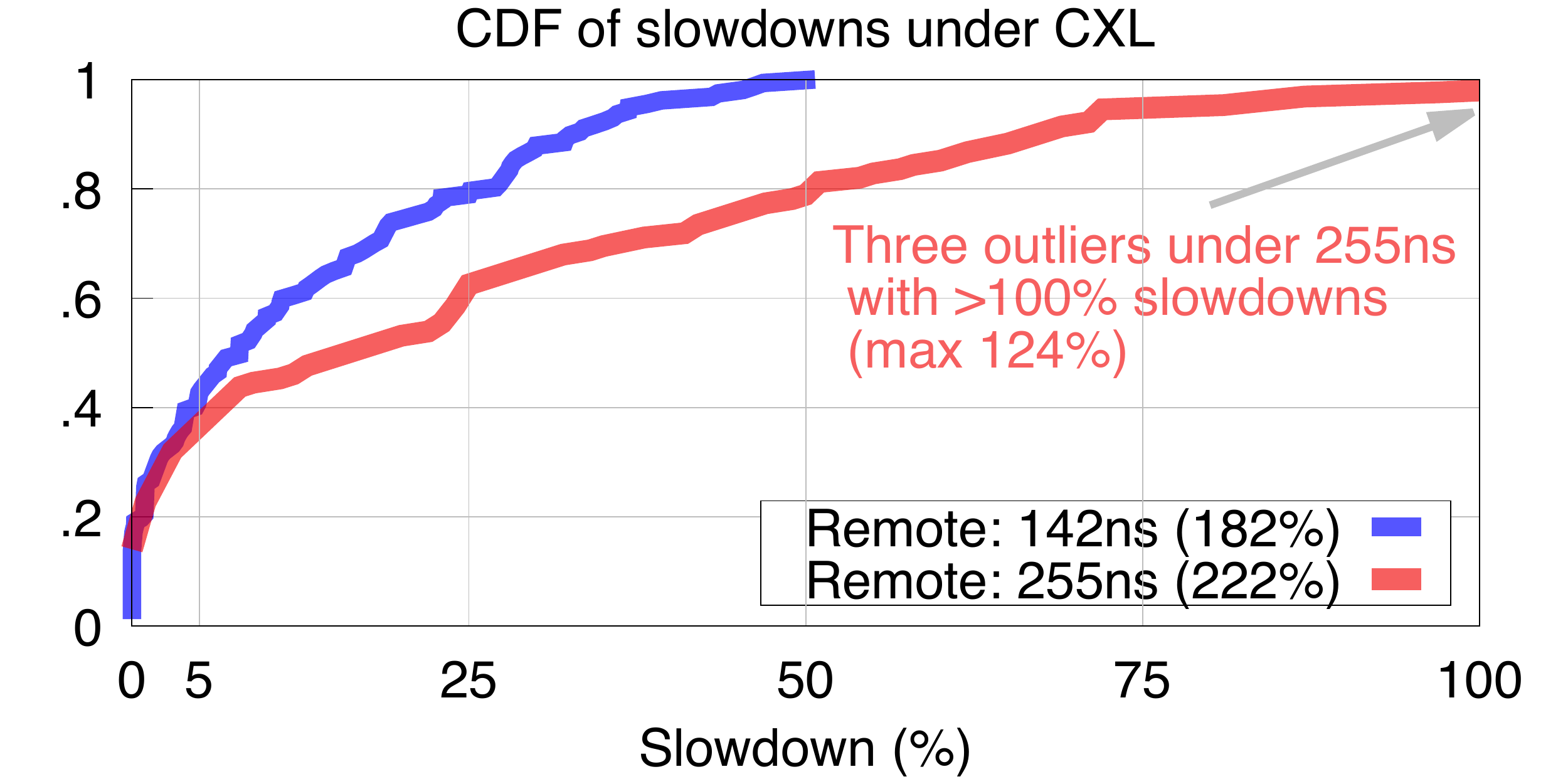}
}


    \mycaption{fig-cxl-cdf}{CDF of slowdowns (\sec\ref{sec-mot-cxl})}{\newtxt{Higher remote latency (red) only slightly affects the head of the distribution (workloads with less than 5\% slowdown). The body and tail of the distribution see significantly higher slowdowns.}}{}

\end{figure}

\myparagraph{Memory stranding} Figure~\ref{fig-stranding-bars}a shows the
daily average amount of stranded
DRAM across clusters, bucketed by the percentage of scheduled CPU cores.
In clusters where 75\% of CPU cores are scheduled for VMs, 6\%
of memory is stranded. This grows to over 10\% when $\sim$85\% of CPU
cores are allocated to VMs. This makes sense since stranding is an
artifact of highly utilized nodes, which correlates with highly utilized
clusters.
Outliers are shown by the error bars, representing 5\th\ and 95\th\
percentiles. At 95\th, stranding reaches 25\% during high utilization
periods. Individual outliers even reach 30\% stranding.

Figure~\ref{fig-stranding-bars}b shows stranding over time across 8
racks. A workload change (around day 36)
suddenly increased stranding significantly.  Furthermore, stranding can
affect many racks concurrently (\eg, racks 2, 4--7) and it is generally
hard to predict which clusters\slash racks will have stranded memory.

\myparagraph{NUMA spanning} Many VMs are small and can fit on a
single socket.  On two-socket systems, the hypervisor at \azure seeks to
schedule such that VMs fit entirely (cores and memory) on a single NUMA node.  In
rare cases, we see \emph{NUMA spanning} where a VM has all of its
cores on one socket and a small amount of memory from
another socket.
We find that spanning occurs for about 2-3\% of VMs and fewer than
1\% of memory pages, on average.

\myparagraph{Savings from pooling}
\azure currently does not pool memory.  However, by analyzing
its VM-to-server traces, we can estimate the amount of DRAM that could
be saved via pooling.  Figure~\ref{fig-stranding-pooling} presents
average reductions from pooling DRAM when VMs are scheduled with a
fixed percentage of either 10\%, 30\%, or 50\% of pool DRAM. The pool
size refers to the number of sockets that can access the same DRAM
pool. As the pool size increases, the figure shows that required
overall DRAM decreases.  However, this effect diminishes for larger
pools.  For example, with a fixed 50\% pool DRAM, a
pool with 32 sockets saves 12\% of DRAM while a pool with 64 sockets
saves 13\% of DRAM.
Note that
allocating a fixed 50\% of memory to pool DRAM leads to significant
performance loss compared to socket-local DRAM (\sec\ref{sec-eval}).
\sys overcomes this challenge with multiple techniques (\sec\ref{sec-des}).

\myparagraph{Summary and implications}  From this analysis, we draw a
few important observations and implications for \sys:

\begin{itemize}

\item We observe 3-27\% of stranded memory in production at the 95\th\ percentile, with
    some outliers at 36\%.

\item Almost all VMs fit into one NUMA node.

\item Pooling memory across 16-32 sockets can reduce cluster memory
    demand by 10\%.  This suggests that memory pooling can
        produce significant cost reductions but assumes that a high
        percentage of DRAM can be allocated on memory pools.  When
        implementing DRAM pools with cross-NUMA latencies, providers
        must carefully mitigate potential performance impacts. \border{}

\end{itemize}

\subsection{VM Memory Usage at \azure}
\label{sec-utilization-frigid}

We use \sys's telemetry on opaque VMs (\sec\ref{sec:design:systemsw}) to characterize the percentage of
untouched memory across our cloud clusters.
Generally, we find that while VM memory usage varies across clusters,
all clusters have a significant fraction of VMs with untouched memory.
Overall, the 50\th\ percentile is 50\% untouched memory.



\myparagraph{Summary and implications}  From this analysis, we draw key
observations and implications for \sys:

\begin{itemize}

\item VM memory usage varies widely.
\item In the cluster with the least amount of untouched memory, still over
    50\% of VMs have more than 20\% untouched memory.  Thus, there is
        plenty of untouched memory that can be disaggregated at no
        performance penalty.
\item The challenges are (1) predicting how much untouched
        memory a VM is likely to have and (2) confining the VM's accesses
        to local memory. \sys addresses both. \border{}

\end{itemize}


\subsection{Workload Sensitivity to Memory Latency}
\label{sec-mot-cxl}

\newtxt{To characterize the performance impact of CXL latency for typical workloads in \azure's datacenters,
we evaluate \numTotalApps\ workloads under two scenarios of emulated CXL access latencies: 182\% and 222\% increase in memory latency, respectively.
We then compare the workload performance to NUMA-local memory placement.
Experimental details are in \sec\ref{sec-eval-setup}.
Figures~\ref{fig-cxl} and~\ref{fig-cxl-cdf} show workload slowdowns relative to NUMA-local performance for both scenarios.}


\begin{figure}[t!]
  \centering
\includegraphics[width=0.99\columnwidth]{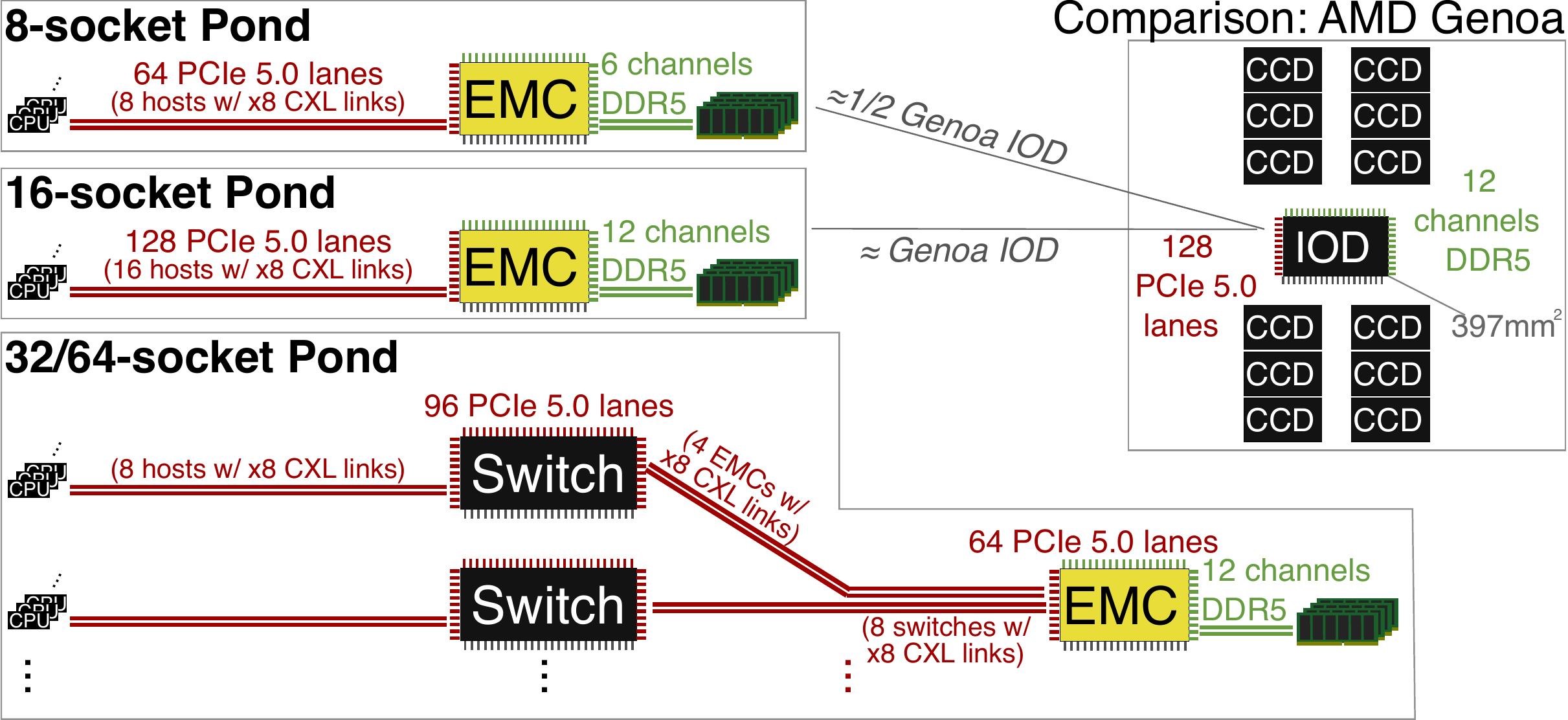}
\vminfive
\mycaption{fig-contactcount-revised}{External memory controller (EMC) (\sec\ref{sec:design:hw})}{\newtxt{The EMC is multi-headed which allows connecting multiple CXL hosts and DDR5 DIMMs. A 16-socket \sys requires 128 PCIe 5.0 lanes and 12 DDR5 channels, which is comparable to the IO-die (IOD) on AMD Genoa~\cite{amdgenoa,genoa2021leak}. Larger \sys configurations combine a switch with the multi-headed EMC.}}
\end{figure}

\newtxt{Under a 182\% increase in memory latency, we find that 26\% of the \numTotalApps\ workloads
experience less than 1\% slowdown under CXL.
An additional 17\% of workloads see less than 5\% slowdowns.
At the same time, some workloads are severely affected with 21\% of the workloads facing $>$25\% slowdowns.}

\newtxt{Different workload classes are affected differently, \eg, GAPBS (graph processing) workloads generally see higher slowdowns.
However, the variability within each workload class is typically much higher than across workload classes.
For example, within GAPBS even the same graph kernel reacts very differently to CXL latency, based on different graph datasets.
Overall, every workload class has at least one workload with less than 5\% slowdown and one workload with more than 25\% slowdown (except SPLASH2x).}

\newtxt{\azure's proprietary workloads are less impacted than the overall workload set.
Of the 13 production workloads, 6 do not see noticeable impact ($<$1\%);
2 see $\sim$5\% slowdown;
and the remaining half are impacted by 10--28\%.
This is in part because these production workloads are NUMA-aware and often include data placement optimizations.}

\newtxt{Under a 222\% increase in memory latency, we find that 23\% of the \numTotalApps\ workloads
experience less than 1\% slowdown under CXL.
An additional 14\% of workloads see less than 5\% slowdowns.
More than 37\% of workloads face $>$25\% slowdowns.
Generally, we find that higher latency magnifies the effects seen under lower latency: workloads performing well under 182\% latency also tend to perform well under 222\% latency; workloads severely affected by 182\% are even more affected by 222\%.}


\myparagraph{Summary and implications}
While the performance of some workloads is insensitive to disaggregated memory latency, some are heavily impacted.
This motivates our design decision to include socket-local DRAM alongside pool DRAM to mitigate CXL latency impact for those latency-sensitive workloads.
Memory pooling solutions can be effective if they're are effective at identifiying sensitive workloads.


\section{\sys Design}
\label{sec-des}


Our measurements and observations at \azure (\sec\ref{sec-bg}--\ref{sec:measurements}) lead us to define the following design goals.
\begin{description}
\item[G1] Performance comparable to NUMA-local DRAM
\item[G2] Compatibility with virtualization accelerators
\item[G3] Compatibility with opaque VMs and unchanged guest OSes\slash applications
\item[G4] Low host resource overhead
\end{description}

To quantify (G1), we define a \emph{performance degradation margin} (\pdm) for a given workload as the allowable slowdown relative to running the workload entirely on NUMA-local DRAM.
\sys seeks to achieve a configurable \pdm, \eg, 1\%, for a configurable tail-percentage (\tp) of VMs, \eg, 98\% (\sec\ref{sec-strand}).
To achieve this high performance, \sys uses a small but fast CXL pool (\sec\ref{sec:design:hw}).
As \sys's memory savings come from pooling instead of oversubscription, \sys must minimize pool fragmentation and wastage in its system software layer (\sec\ref{sec:design:systemsw}).
To achieve (G2), \sys preallocates local and pool memory at VM start.
\sys decides this allocation in its allocation, performance monitoring, and mitigation pipeline (\sec\ref{sec:design:distributedsw}).
This pipeline uses novel prediction models to achieve the \pdm (\sec\ref{sec:design:ml}).
Finally, \sys overcomes VM-opaqueness (G3) and host-overheads (G4) using lightweight hardware counter telemetry.



\begin{figure}[t!]
  \centering
\includegraphics[width=\columnwidth]{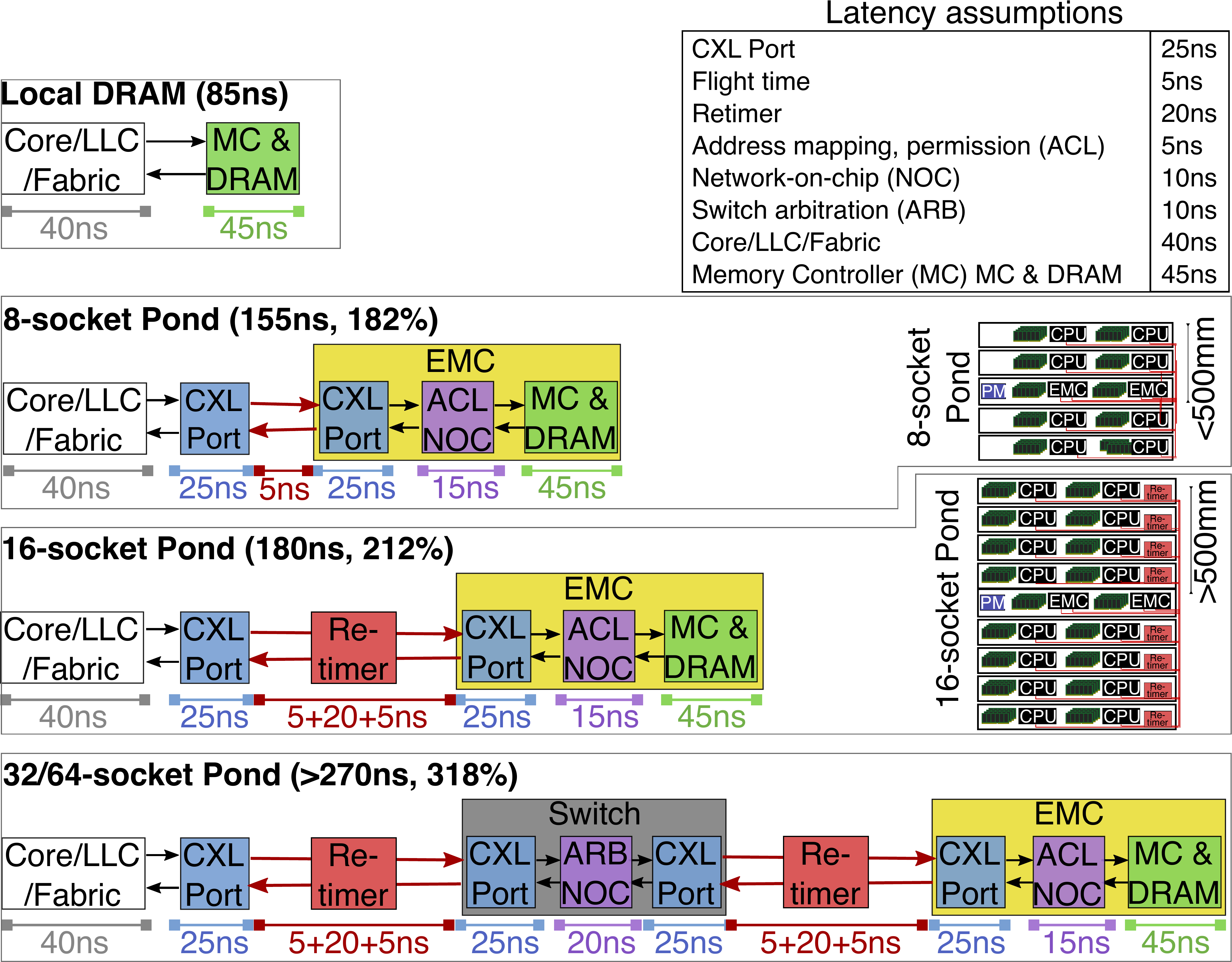}
  \vminten
    \mycaption{fig-analysis-overheads}{Pool size and latency tradeoffs (\sec\ref{sec:design:hw})}{\newtxt{Small \sys pools of 8-16 sockets add only 75-90ns relative to NUMA-local DRAM.
  Latency increases for larger pools that require retimers and a switch.}}
\end{figure}

\subsection{Hardware Layer}\label{sec:design:hw}



\newtxt{Hosts within a \sys pool have separate cache coherency domains and run separate hypervisors.
\sys uses an ownership model where pool memory is explicitly moved among hosts.
A new external memory controller (EMC) ASIC implements the pool using multiple DDR5 channels accessed through a collection of CXL ports running at PCIe 5 speeds.}


\myparagraph{\newtxt{EMC memory management}}
\newtxt{The EMC offers multiple CXL ports and appears to each host as a single logical memory device~\cite{cxl2spec.web20,cxl2whitepaper.web21}.
In CXL 3.0~\cite{debendra2022fms,cxl3spec}, this configuration is standardized as multi-headed device (MHD)~\cite[\S2.5]{cxl3spec}.
The EMC exposes its entire capacity on each port (\eg, to hosts) via a Host-managed Device Memory (HDM) decoder.
Hosts program each EMC's address range but treat them initially as offline.
\sys dynamically assigns memory at the granularity of 1GB memory slices.
Each slice is assigned to at most one host at a given time and hosts are explicitly notified about changes (\S\ref{sec:design:systemsw}).
Tracking 1024 slices (1TB) and 64 hosts (6 bits) requires 768B of EMC state.
The EMC implements dynamic slice assignment by checking permission of each memory access, \ie, whether requestor and owner of the cacheline's slice match.
Disallowed accesses result in fatal memory errors.}

\begin{figure}[t!]
  \centering
\includegraphics[width=\columnwidth]{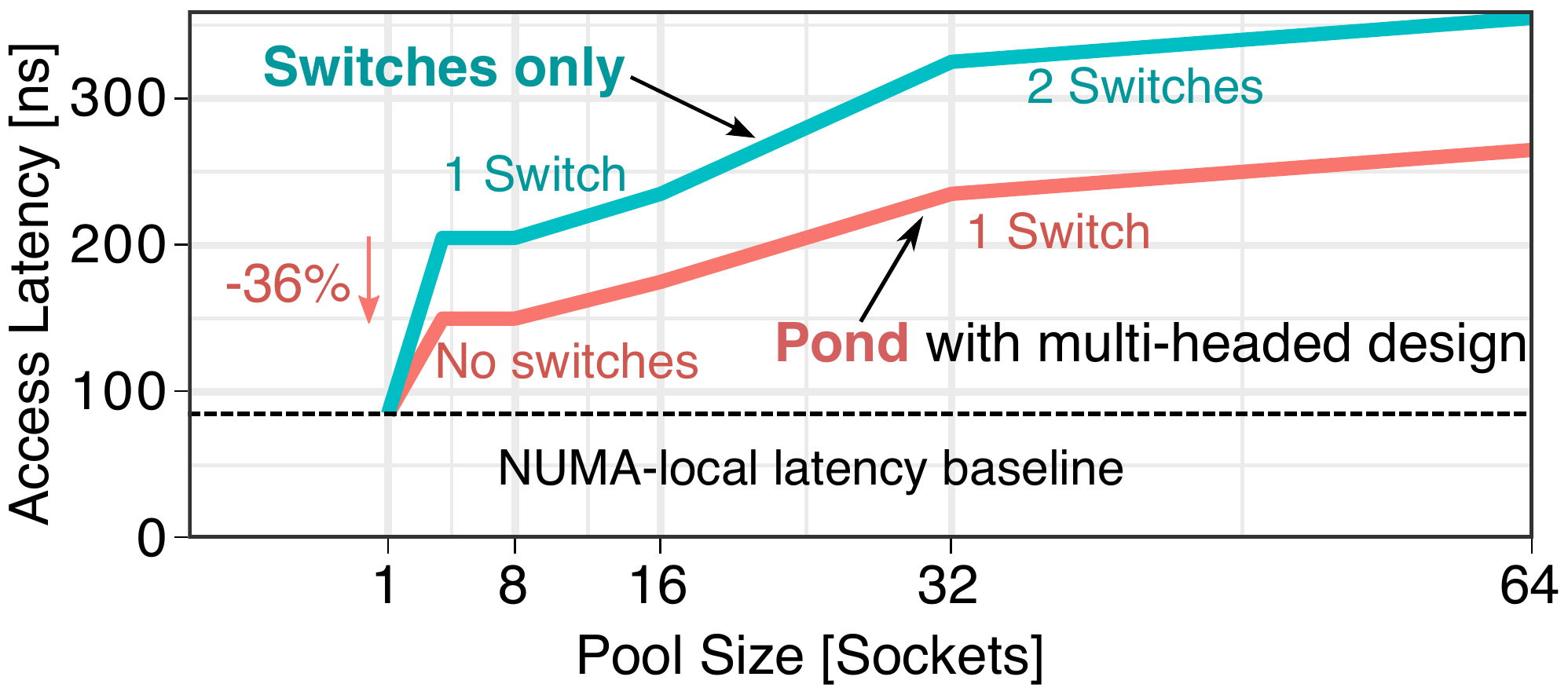}
\vminfifteen
    \mycaption{fig-poollatency}{Pool access latency comparison (\sec\ref{sec:design:hw})}{\newtxt{\sys reduces latencies by 1/3 compared to switch-only designs.}}
    \vfive
\end{figure}

{\myparagraph{\newtxt{EMC ASIC design}}
\newtxt{The EMC offers multiple \tms{8}-CXL ports, which communicate with DDR5 memory controllers (MC) via an on-chip network (NOC).
The MCs must offer the same reliability, availability, and serviceability capabilities~\cite{intelras,amdras} as server-grade memory controllers including memory error correction, management, and isolation.
A key design parameter of \sys's EMC is the pool size, which defines the number of CPU sockets able to use pool memory.
We first observe that the EMC's IO, (De)Serializer, and MC requirements resemble AMD Genoa's $397mm^2$ IO-die (IOD)~\cite{amdgenoa,genoa2021leak}.
Figure~\ref{fig-contactcount-revised} shows that EMC requirements for a 16-socket \sys parallel the IOD's requirements, with a small 8-socket \sys paralleling half an IOD.
Thus, up to 16-sockets can directly connect to an EMC.
Pool sizes of 32-64 would combine CXL switches with \sys's multi-headed EMC.
The optimal design point balances the potential pool savings for larger pool sizes (\S\ref{sec-eval}) with the added cost of larger EMCs and switches.}


\myparagraph{\newtxt{EMC Latency}}
\newtxt{
While latency is affected by propagation delays, it is dominated by CXL port latency, and any use of CXL retimers and CXL switches.
Port latencies are discussed in \sec\ref{sec-bg} and~\cite{debendra2022hoti}.
Retimers are devices used to maintain CXL/PCIe signal integrity over longer distances and add about 10ns of latency in each direction~\cite{microchipretimer,asteraretimer}.
In datacenter conditions, signal integrity simulations~\cite{cxlretimers} indicate that CXL could require retimers above 500mm.
Switches add at least 70ns of latency due to ports/arbitration/NOC with estimates above 100ns~\cite{cxlswitchlatency}.}

\newtxt{Figure~\ref{fig-analysis-overheads} breaks down \sys's latency for different pool sizes.
Figure~\ref{fig-poollatency} compares \sys's latency to a design that relies only on switches instead of a multi-headed EMC.
We find that \sys reduces latencies by 1/3 with 8-and 16-socket pools adding only 70-90ns relative to NUMA-local DRAM.
In practice, we expect \sys to be deployed primarily with small 8/16-socket pools, given the latency and cost overheads, and diminishing returns of larger pools (\sec\ref{sec:measurements}).
Modern CPUs can connect to multiple EMCs which allows scaling to meet bandwidth and capacity goals for different clusters.}

\def \hmina {\hspace{0.05in}}
\def \hminb {\hspace{-0.2in}}

\begin{figure}[t!]
\centering
\hmina
\includegraphics[width=\columnwidth]{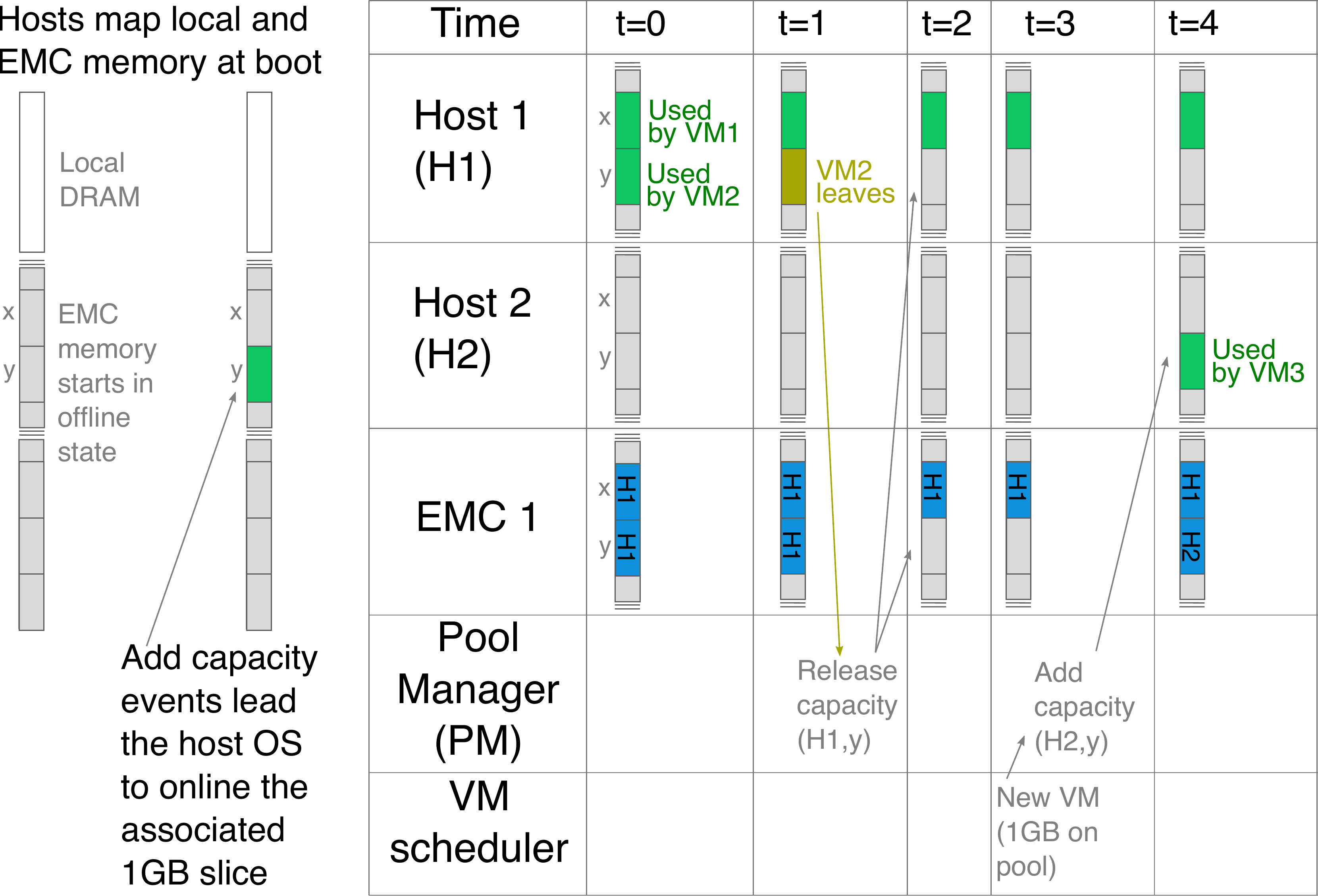}
%
\vminfive
    \mycaption{fig-poolmng}{Pool management example (\sec\ref{sec:design:systemsw})}{
\newtxt{\sys assigns pool memory to at most one host at a time.
This example shows \sys's asynchronous memory release strategy which engages when a VM departs ($t$=1 and $t$=2).
During VM scheduling, memory is added to the corresponding host before the VM starts ($t$=3 and $t$=4).}}
\vten
\end{figure}

\subsection{System Software Layer}
\label{sec:design:systemsw}

\sys's system software involves multiple components.

\myparagraph{Pool memory ownership}
Pool management involves assigning \sys's memory slices to hosts and reclaiming them for the pool (Figure~\ref{fig-poolmng}).
It involves 1) implementing the control paths for pool-level memory assignment and 2) preventing pool memory fragmentation.

\newtxt{Hosts discover local and pool capacity through CXL device discovery and map them to their address space. 
Once mapped, the pool address range is marked hot-pluggable and ``not enabled.''}
\newtxt{Slice assignment is controlled at runtime via a Pool Manager (PM) that is colocated on the same blade as the EMCs (Figure~\ref{fig-analysis-overheads}).
In \sys's current design, the PM is connected to EMCs and CPU sockets via a low-power management bus
(\eg,~\cite{i3c}).
To allocate pool memory, the Pool Manager triggers two types of interrupts at the EMC and host driver.
\ts{Add\_capacity(host, slice)} interrupts the host driver which reads the address range to be hot-plugged.
The driver then communicates with the OS memory manager to bring the memory online.
The EMC adds the host id to its permission table at the slice offset.
\ts{Release\_capacity(host, slice)} works similarly by offlining the slice on the host and resetting the slice's permission table entry on the EMC.
An alternative to this design would be inband-communication using the Dynamic Capacity Device (DCD) feature in CXL 3.0~\cite[\S9.13]{cxl3spec}.
This change would maintain the same functionality for \sys.}


\sys must avoid fragmenting its online pool memory as the contiguous 1GB address range must be free before it can be offlined for reassignment to another host.
Pool memory is allocated to VMs in 1GB-aligned increments (\sec\ref{sec:design:distributedsw}).
While this prevents fragmentation due to VM starts and completions, our experience has shown that host agents and drivers can allocate pool memory and cause fragmentation.
\sys thus uses a special-purpose memory partition that is only available to the hypervisor.
Host agents and drivers allocate memory in host-local memory partition, which effectively contains fragmentation.

With these optimizations, offlining 1GB slices empirically takes 10-100 milliseconds\slash GB.
Onlining memory is near instantaneous with microseconds\slash GB.
These observations are reflected in \sys's asynchronous release strategy (\sec\ref{sec:design:distributedsw}).


\myparagraph{Failure management}
Hosts only interleave across local memory.
This minimizes the EMCs' blast radius and facilitate memory hot-plugging.
EMC failures affect only VMs with memory on that EMC, while VMs with memory on other EMCs continue normally.
CPU\slash host failures are isolated and associated pool memory is reallocated to other
hosts.
Pool Manager failures prevent reallocating pool memory but do not affect the datapath.


\def \vmina {\vspace{-0.1in}}
\def \vminb {\vspace{0.05in}}

\begin{figure}[t!]
\centering
\includegraphics[width=0.99\columnwidth]{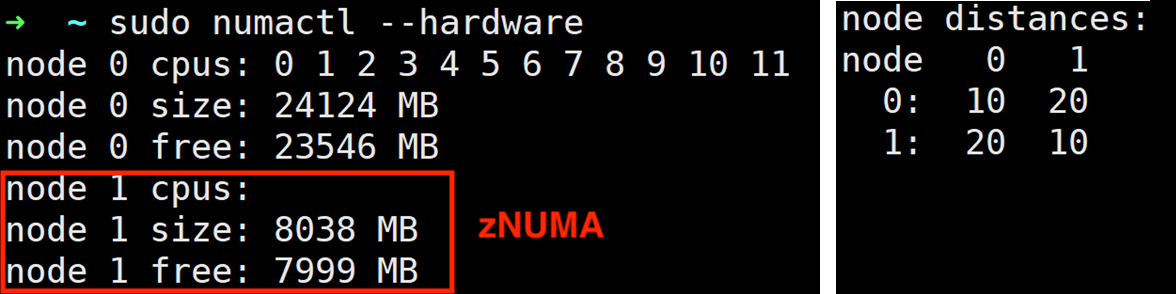}


    \mycaption{fig-vnuma}{zNUMA (\sec\ref{sec:design:systemsw})}{zNUMA
seen from a Linux VM.}

\vfive

\end{figure}

\myparagraph{Exposing pool memory to VMs}
VMs that use both NUMA-local and pool memory see pool memory as a \cvn node.
The hypervisor creates a \cvn node by adding a memory block (\texttt{node\_memblk}) without
an entry in the \texttt{node\_cpuid} in the
\ts{SLIT\slash SRAT} tables~\cite{acpi}.
We later show the guest-OS preferentially allocates memory from the local NUMA node before going to \cvn (\sec\ref{sec-eval}).
Thus, if \cvn is sized to the amount of untouched memory, it is never going to be used.
Figure~\ref{fig-vnuma} shows the view of a Linux VM which includes
the correct latency in the NUMA distance
matrix (\texttt{numa\_slit}).
This facilitates guest-OS NUMA-aware memory management~\cite{nimblepage.asplos19, autonuma.web19} for the rare case that the \cvn is used (\sec\ref{sec:design:ml}).

\def \hmina {\hspace{-0.1in}}
\def \hminb {\hspace{-0.2in}}

\begin{figure}[t!]
\centering
\includegraphics[width=0.98\columnwidth]{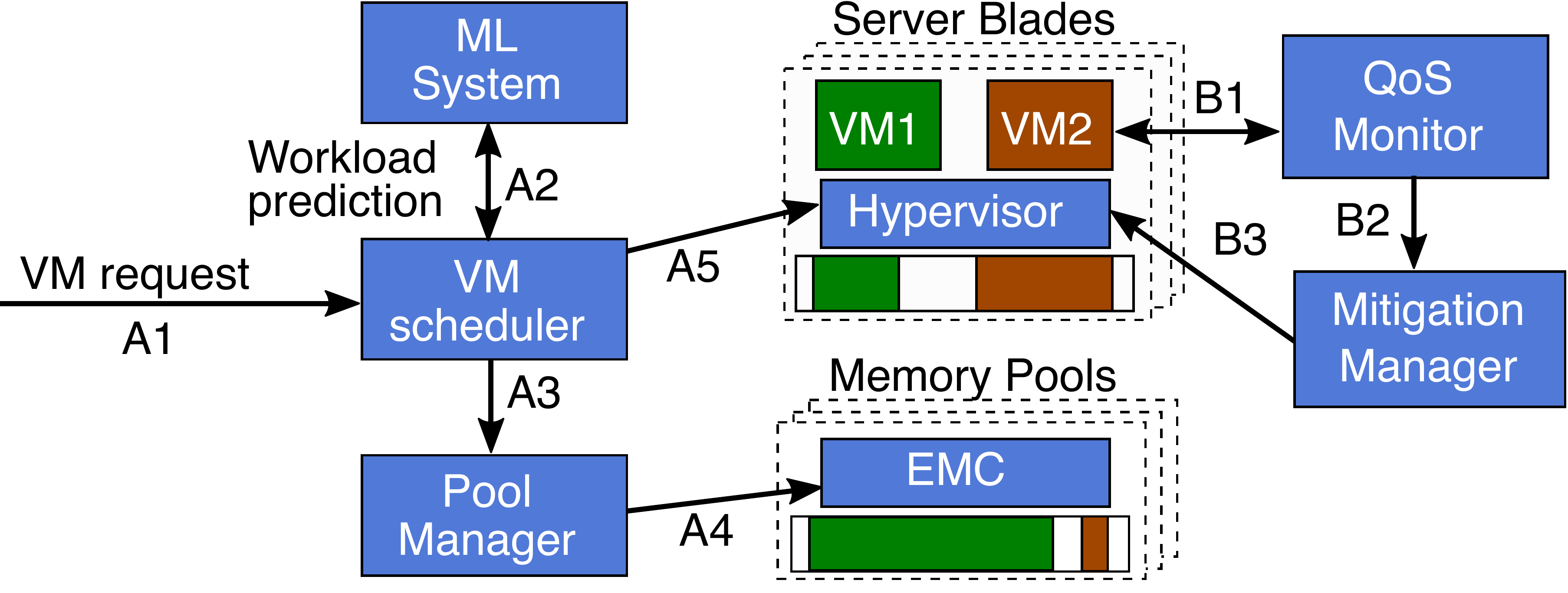}
\vtri
\mycaption{fig-workflow}{\sys control plane workflow
(\sec\ref{sec:design:distributedsw})}{
{\bf A)}~The VM scheduler uses ML-based
predictions that identify latency-sensitive VMs and their likely
amount of untouched memory to decide on VM placement (see Figure~\ref{fig-premodels}).
{\bf B)}
The monitoring pipeline reconfigures VMs if quality-of-service (QoS) not
met.}
\vten

\end{figure}

\myparagraph{Reconfiguration of memory allocation}
To remain compatible with ({\bf G2}), local and pool memory mapping generally remain static during a VM's lifetime.
There are two exceptions that are implemented today.
When live-migrating a VM or when remapping a page with a memory fault, the hypervisor temporarily disables virtualization acceleration
and the VM falls back to a slower I/O path~\cite{ruprecht2018vm}.
Both events are quick and transient and typically only happen once during a VM's lifetime.
We implement a third variant which allows \sys a one-time correction to a suboptimal memory allocation.
Specifically, if the host has local memory available, \sys disables the accelerator, copies all of the VM's memory to local memory, and enables the accelerator again.
This takes about 50ms for every GB of pool memory that \sys allocated to the VM.

\myparagraph{Telemetry for opaque VMs}
\sys requires two types of telemetry for VMs.
First, we use the core-performance-measurement-unit (PMU) to gather hardware counters related to memory performance.
Specifically, we use the top-down-method for analysis
(TMA)~\cite{topdownanalysis.ispass14,tmam.web21}.
TMA characterizes how the core pipeline slots are used.
For example, we use the ``memory-bound'' metric, which
is defined as pipeline stalls due to memory loads and stores.
Figure~\ref{ml-training1} lists these metrics.
While TMA was developed for Intel, its relevant parts are available on AMD and
ARM as well~\cite{jarus2016top}.
We modify \azure's production hypervisor to associate these metrics with individual
VMs (\sec\ref{sec-impl}) and record VM counter samples in a distributed database.
All our core-PMU-metrics use simple counters and induce negligible overhead (unlike event-based sampling~\cite{akiyama2017quantitative,hemem.sosp21}).

Second, we use hypervisor telemetry to track a VM's untouched pages.
We use an existing hypervisor counter that tracks guest-committed memory, which overestimates used memory.
This counter is available for 98\% of \azure VMs.
We also scan access bits in the hypervisor page table (\sec\ref{sec-impl}).
Since we only seek untouched pages, frequently access bits reset is not required. This minimizes overhead.

\subsection{Distributed Control Plane Layer}
\label{sec:design:distributedsw}

Figure~\ref{fig-workflow} shows the two tasks performed by \sys's control plane:
(A) predictions to allocate memory during VM scheduling
and (B) QoS monitoring and resolution.

\myparagraph{Predictions and VM scheduling (A)}
\sys uses ML-based prediction models (\sec\ref{sec:design:ml})
to decide how much pool memory
to allocate for a VM during scheduling.
After a VM request arrives (A1), the scheduler queries the
distributed ML serving system (A2) for a prediction on how much local memory
to allocate for the VM. The scheduler
then informs the Pool Manager about the target host and associated pool memory
needs (A3).
The Pool Manager triggers a memory onlining workflow using the configuration
bus to the EMCs and host (A4).
Memory onlining is fast enough to not block a VM's start time (\sec\ref{sec:design:systemsw}).
The scheduler informs the hypervisor
to start the VM on a \cvn node matching the onlined memory amount.

Memory offlining is slow and cannot happen on the critical path of VM starts (\sec\ref{sec:design:systemsw}).
\sys resolves this by always keeping a buffer of unallocated pool memory.
This buffer is replenished when VMs terminate and hosts asynchronously release associated slices.

\begin{figure}[t!]
\centering
\includegraphics[width=0.75\columnwidth]{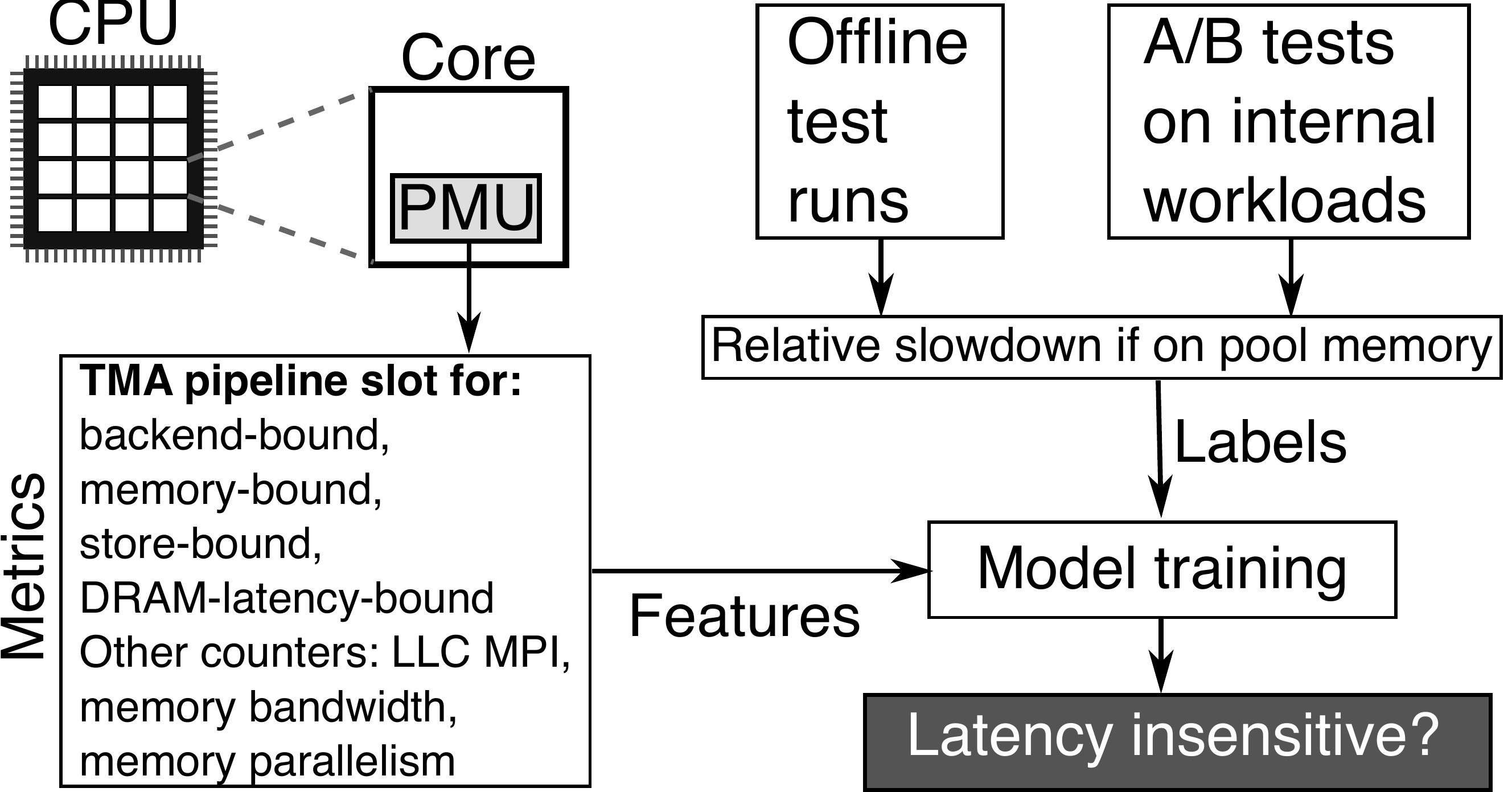}
\vtri
\mycaption{ml-training1}{Metrics and training of the latency
    insensitive model (\sec\ref{sec:design:systemsw})}{This model uses metrics from the
core's performance-measurement-unit (PMU). It is trained with labels
gathered from offline runs and internal workloads.}
\vten
\end{figure}

\myparagraph{QoS monitoring (B)} \sys
continuously inspects the performance of all running VMs via its QoS monitor.
The monitor queries hypervisor and hardware performance counters (B1) and
uses an ML model of latency sensitivity (\sec\ref{sec:design:ml})
to decide whether the VM's performance impact exceeds the \pdm.
In this case, the monitor asks its mitigation manager (B2) to
trigger a memory reconfiguration (\sec\ref{sec:design:systemsw}) through the hypervisor (B3).
After this reconfiguration, the VM uses only local memory.

\subsection{Prediction Models}
\label{sec:design:ml}

\sys's VM scheduling (A) and QoS monitoring (B) algorithms rely on two
prediction models (in Figure~\ref{fig-premodels}).

\myparagraph{Predictions for VM scheduling (A)} For scheduling, we first
check if we can correlate a workload history with the VM requested.
This works by checking if there have been previous VMs with the same
metadata as the request VM, \eg, the customer-id, VM type, and location.
This is based on the observation that
VMs from the same customer tend to exhibit similar
behavior~\cite{resourcecentral.sosp17}.

If we have prior workload history, we make a prediction on whether this
VM is likely to be memory latency insensitive, \ie, its performance would be
within the \pdm while using only pool memory. (Model details appear below.)
Latency-insensitive VMs are allocated entirely on pool DRAM.

\def \hmina {\hspace{-0.1in}}
\def \hminb {\hspace{-0.2in}}

\begin{figure}[t!]
\centering
\includegraphics[width=\columnwidth]{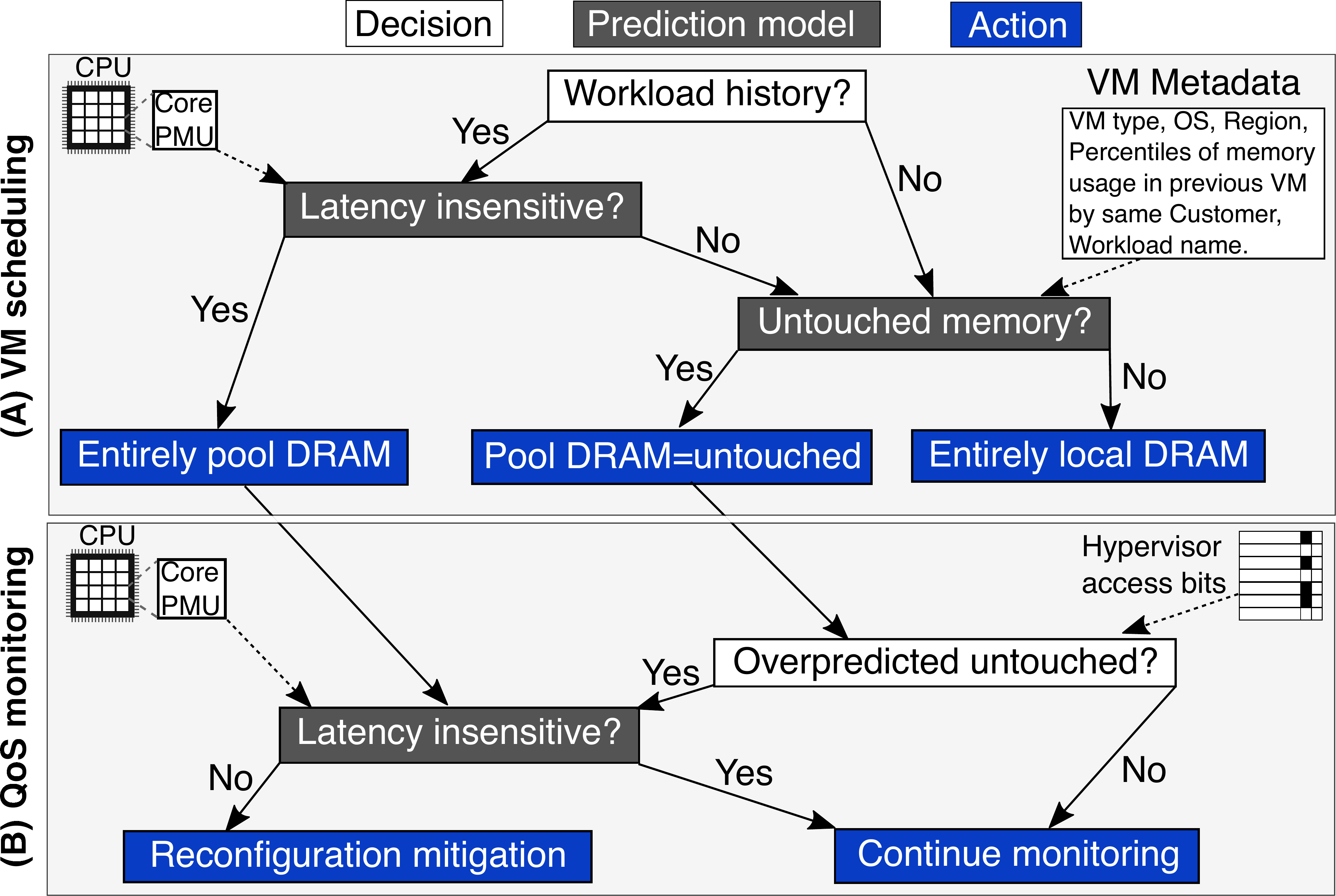}
\vminfive
\mycaption{fig-premodels}{\sys prediction models (\sec\ref{sec:design:ml})}{
\sys's two prediction models (dark grey) rely on telemetry (blue boxes) that is available for all VM types, including third-party opaque VMs.}
\vten

\end{figure}

If the VM has no workload history or is predicted to be latency-sensitive,
we predict untouched memory ($\texttt{UM}$) over its
lifetime.  Interestingly, $\texttt{UM}$ predictions with only generic VM metadata such as customer history, VM type, guest OS,
and location are accurate (\sec\ref{sec-eval}).  VMs without untouched memory ($\texttt{UM}=0$) are allocated
entirely with local DRAM.  VMs with a $\texttt{UM}>0$ are allocated with a rounded-down GB-aligned percentage of pool
memory and a corresponding \cvn node;
the remaining memory is allocated on local DRAM.

If we underpredict $\texttt{UM}$, the VM will not touch the slower pool
memory as the guest OS prioritizes allocating local DRAM.
If we overpredict $\texttt{UM}$, we rely on the QoS monitor for mitigation.
Importantly, \sys always keeps a VM's memory mapped in hypervisor page
tables at all times. This means that even if our predictions happen to
be incorrect, performance does not fall off a cliff.

\myparagraph{QoS monitoring (B)}
For \cvn VMs, \sys monitors if it overpredicted
the amount of untouched memory during scheduling.
For pool-backed VMs and \cvn VMs with
less untouched memory than predicted, we use the sensitivity model to
determine whether the VM workload is suffering excessive performance loss.
If not, the QoS monitor initiates a live VM migration to a configuration
allocated entirely on local DRAM.

\myparagraph{Model details}
\sys's two ML prediction models consume telemetry that is available for opaque VMs from \sys's system software layer
(\sec\ref{sec:design:systemsw}).
Figure~\ref{ml-training1} shows features, labels, and the training procedure for the
latency insensitivity model.
The model uses supervised learning (\S\ref{sec-impl}) with core-PMU metrics as features
and the slowdown of pool memory relative to NUMA-local memory as labels.
\sys gets samples of slowdowns from offline test runs and A\slash B tests of internal
workloads which make their performance numbers available.
These feature-label-pairs are used to retrain the model daily.
As the core-PMU is lightweight (\sec\ref{sec-impl}),
\sys continuously measures core-PMU metrics at VM runtime.
This enable the QoS monitor to react quickly and enables retaining
a history of VMs that have been latency sensitive.

\newtxt{Figure~\ref{ml-training2} shows the inputs and training procedure for the
untouched-memory model.
The model uses supervised learning (details in \S\ref{sec-impl}) with VM metadata as features and the minimum untouched memory over each VM's lifetime as labels.
Its most important feature is a range of percentiles (\eg, 80\th--99\th) of the recorded untouched memory by a customer's VMs in the last week.}

\begin{figure}[t!]
  \centering
\includegraphics[width=1.01\columnwidth]{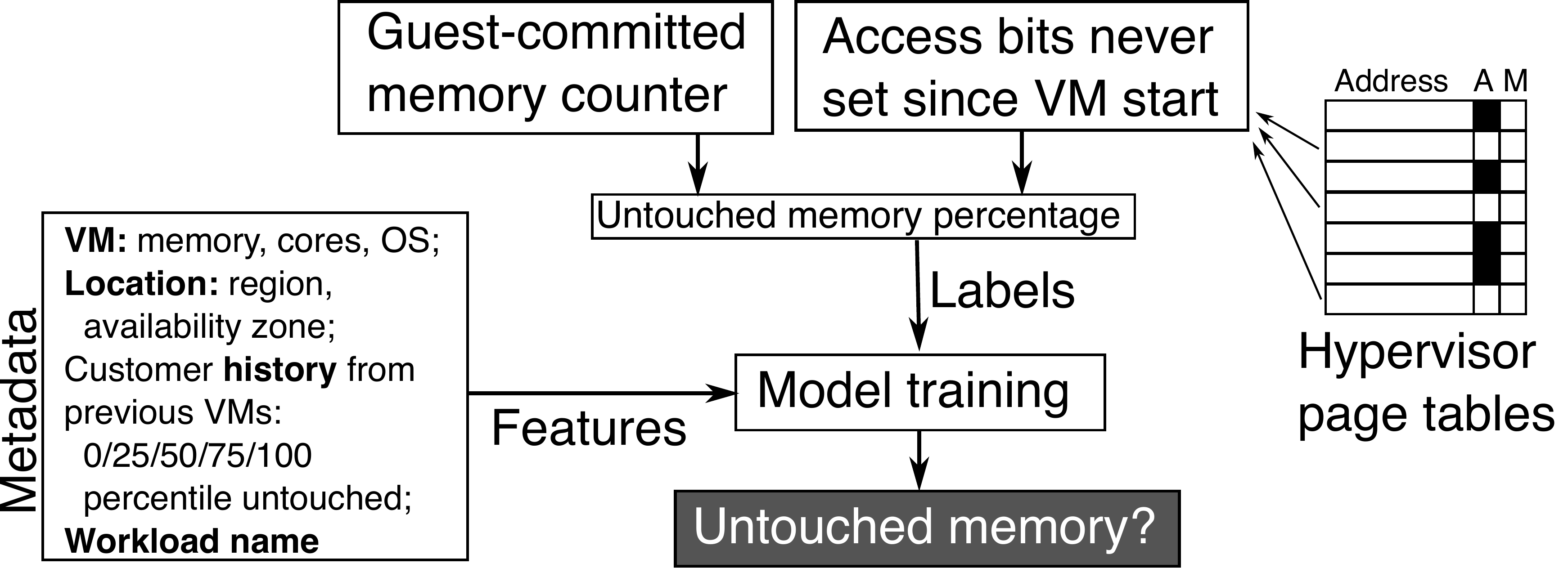}
\vminfive
    \mycaption{ml-training2}{Training of the untouched memory model (\sec\ref{sec:design:ml})}{This model uses VM metadata as features and labels of untouched memory gathered from hypervisor telemetry.}
    \vten
\end{figure}

\myparagraph{Parameterization of prediction models}
\sys's latency insensitivity model is parameterized to stay below a \emph{target rate of false positives} (\texttt{FP}), \ie, workloads it incorrectly specifies as latency insensitive but which are actually sensitive to memory latency.
This parameter enforces a tradeoff as the percentage of workloads that are labeled as latency insensitive (\texttt{LI}) is a function of \texttt{FP}.
For example, a rate of 0.1\% \texttt{FP} may force the model to 5\% of \texttt{LI}.

Similarly, \sys's untouched memory model is parameterized to stay below a \emph{target rate of overpredictions} (\texttt{OP}), \ie, workloads that touch more memory than predicted and thus would use memory pages on the \cvn node.
This parameter enforces a tradeoff as the percentage of untouched memory (\texttt{UM}) is a function of \texttt{OP}.
For example, a rate of 0.1\% \texttt{OP} may force the model to 3\% of \texttt{UM}.

With two models and their respective parameters, \sys needs to decide how to balance \texttt{FP} and \texttt{OP} between the two models.
This balance is done by solving an optimization problem based on the given performance degradation margin (\pdm) and the target percentage of VMs that meet this margin (\tp).
Specifically, \sys seeks to maximize the average amount of memory that is allocated on the CXL pool, which is defined by \texttt{LI} and \texttt{UM}, while keeping the percentage of false positives (\texttt{FP}) and untouched overpredictions (\texttt{OP}) below the \tp.
\begin{align}
\nonumber  \text{maximize} & \quad (\texttt{LI}_\pdm) + (\texttt{UM})\\
\label{eq:optimization}  \text{subject to} & \thinspace (\texttt{FP}_\pdm) + (\texttt{OP}) \leq (100 - \tp)
\end{align}
Note that \tp essentially defines how often the QoS monitor has to engage and initiate memory reconfigurations.

Besides \pdm and \tp, \sys has no other parameters as it automatically solves the optimization problem from Eq.\eqref{eq:optimization}.
The models rely on their respective framework's default hyperparameters (\sec\ref{sec-impl}).

\section{Implementation}
\label{sec-impl}
We implement and evaluate \sys on production servers that emulate pool latency. Pond artifacts are open-sourced at \url{https://github.com/vtess/Pond}.

\myparagraph{System software}
This implementation comprises three parts.
First, we emulate a single-socket system with a CXL pool on a two-socket server by disabling all cores in one socket, while keeping its memory accessible from the other socket. This memory mimics the pool.

Second, we change \azure's hypervisor to instantiate arbitrary \cvn topologies.
We extend the API between the control plane and the host to pass the desired \cvn topology to the hypervisor.

Third, we implement support in \azure's hypervisor for the telemetry required for training \sys's models.
We extend each virtual core's metadata with a copy of its core-PMU state and transfer this state when it gets scheduled on different physical cores.
\sys samples core-PMU counters once per second, which takes 1ms.
We enable access bit scanning in hypervisor page tables.
We scan and reset access bits every 30 minutes, which takes 10s.

\myparagraph{Distributed control plane}
We train our prediction models by aggregating daily telemetry into a central database.
The latency insensitivity model uses a simple random forest (RandomForest) from Scikit-learn~\cite{pedregosa2011scikit} to classify whether a
workload exceeds the \pdm.
The model uses a set of 200 hardware counters as supported by current Intel processors.
The untouched memory model uses a gradient boosted regression model (GBM) from LightGBM~\cite{ke2017lightgbm} and makes a quantile regression prediction with a configurable target percentile.
After exporting to ONNX~\cite{onnx}, the prototype adds the prediction (the size of \cvn) on the VM request path using a custom inference serving system similar to~\cite{olston2017tensorflow,crankshaw2017clipper,microsoft_trace_analysis}.
\azure's VM scheduler incorporates \cvn requests and pool memory as an additional dimension into its bin packing, similar to other cluster schedulers~\cite{hadary2020protean,burns2016borg,delimitrou2013paragon,delimitrou2015tarcil,schwarzkopf2013omega,verma2015large}.

\section{Evaluation}
\label{sec-eval}

Our evaluation addresses the performance of \cvn VMs (\sec\ref{sec:eval:accessbit}, \sec\ref{sec-eval-split}), the accuracy of \sys's prediction models (\sec\ref{sec:eval:models}), and \sys's end-to-end DRAM savings (\sec\ref{eval:e2e}).

\subsection{Experimental Setup}
\label{sec-eval-setup}

We evaluate the performance of our prototype using \numTotalApps\ cloud workloads.
Specifically, our workloads span in-memory databases and KV-stores (Redis~\cite{redis.web21}, VoltDB~\cite{voltdb.web21}, and TPC-H on MySQL~\cite{tpch.web21}), data and graph processing (Spark~\cite{hibench.web21} and GAPBS~\cite{gapbs.corr15}), HPC (SPLASH2x~\cite{parsec3.can16}), CPU and shared-memory benchmarks (SPEC CPU~\cite{speccpu2017.web21} and PARSEC~\cite{parsec.pact08}), and a range of \azure's internal workloads (Proprietary).
Figure~\ref{fig-cxl} overviews these workloads.
We quantify DRAM savings with simulations.

\myparagraph{Prototype setup} We run experiments on production servers at \azure and similarly-configured lab servers.
The production servers use either two Intel Skylake 8157M sockets with each 384GB of DDR4, or two AMD EPYC 7452 sockets with each 512GB of DDR4.
On Intel, we measure 78ns NUMA-local latency and 80GB/s bandwidth and 142ns remote latency and 30GB/s bandwidth (3/4 of a CXL \tms{8} link).
On AMD, we measure 115ns NUMA-local latency and 255ns remote latency.
Our BIOS disables hyper-threading, turbo-boost, and C-states.

We use performance results of VMs entirely backed by NUMA-local DRAM as our \emph{baseline}.
We present \cvn performance as
normalized slowdowns, \ie, the ratio to the baseline. Performance metrics are workload specific, \eg, job runtime, throughput and tail latency, \etc

Each experiment involves running the application with one of 7 \cvn sizes (as percentages of the workload's memory footprint in Figure \ref{fig-split-cdf}).
With at least three repetitions of each run and \numTotalApps\ workloads, our evaluation spans more than 3,500 experiments and 10,000 machine hours.
Most experiments used lab servers; we spot check outliers on production servers.

\myparagraph{Simulations}
Our simulations are based on traces of production VM requests and their placement on servers.
The traces are from randomly selected 100 clusters across 34 datacenters globally over 75 days.

The simulator implements different memory allocation policies and tracks each server and each pool's memory capacity at second accuracy.
Generally, the simulator schedules VMs on the same nodes as in the trace and changes their memory allocation to match the policy.
For rare cases where a VM does not fit on a server, \eg, due to insufficient pool memory, the simulator moves the VMs to another server.

\myparagraph{Model evaluation}
We evaluate our model with production resource logs.
About 80\% of VMs have sufficient history to make a sensitivity prediction.
Our deployment does not report each workload's perceived performance (opaque VMs).
We thus evaluate latency sensitivity model based on our \numTotalApps\ workloads.

\def \hmina {\hspace{0.00in}}
\def \hminb {\hspace*{-0.05in}}

\begin{figure}[t!]
\hmina
\begin{center}

\begin{minipage}[t]{\columnwidth}
\hspace{-1mm}
\includegraphics[width=.9\columnwidth]{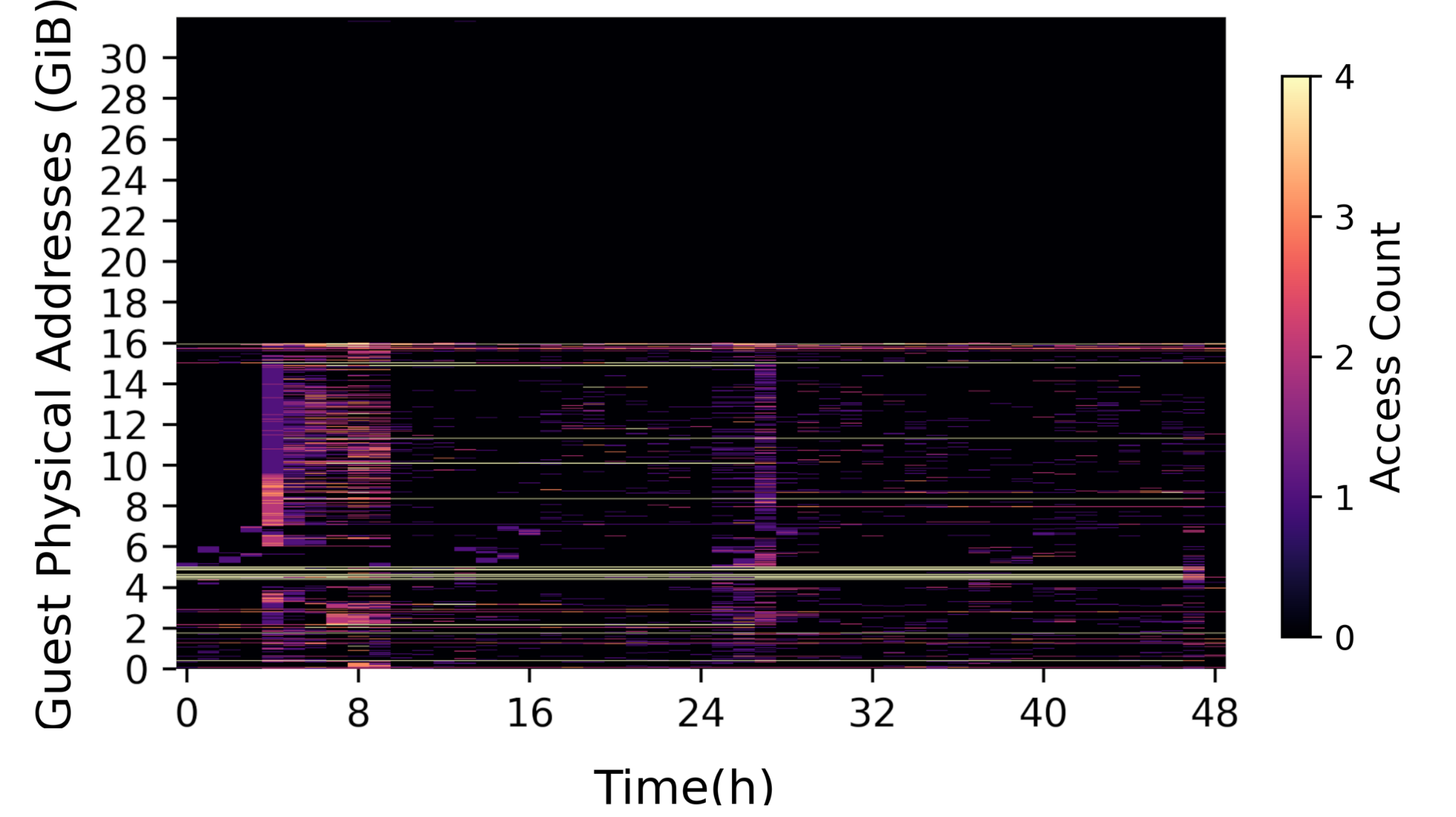}
\end{minipage}

\vspace*{16pt}

\begin{minipage}[t]{\columnwidth}
\setlength{\tabcolsep}{4pt}

\vspace{-8mm}
\begin{center}
\small
\begin{tabular}{l|>{\raggedleft\arraybackslash}p{30mm}}
Workloads & Traffic to \cvn \\
\hline
Video & 0.25\%\\
Database & 0.06\% \\
KV store & 0.11\% \\
Analytics & 0.38\%
\end{tabular}
\end{center}
\vspace{-6mm}
\end{minipage}

\vspace*{12pt}

\mycaption{fig-accessbit}{Effectiveness of \cvn (\sec\ref{sec:eval:accessbit})}{Latency sensitive workloads get a local vNUMA node large enough to cover the workload's footprint. \cvn nodes holds the VM's remaining memory on \sys CXL pool.
Access bit scans, \eg, for Video (right), show that this configuration indeed minimizes traffic to the \cvn node.
}

\vminten
\end{center}
\end{figure}

\subsection{\cvn VMs on Production Nodes}
\label{sec:eval:accessbit}

We perform a small-scale experiment on \azure production nodes to validate \cvn VMs.
The experiment evaluates four internal workloads:
an audio/video conferencing application,
a database service, a key-value store, and a business analytics service.
To see the effectiveness of \cvn, we assume a correct prediction of untouched memory, \ie, the local footprint fits into the VM's local vNUMA node.
Figure~\ref{fig-accessbit} shows access bit scans over 48 hours from the video workload and a table that shows the traffic to the \cvn node for the four workloads.

\myfinding{}
We find that \cvn nodes are effective at containing the memory access to the local vNUMA node.
A small fraction of accesses goes to the \cvn node.
We suspect that this is in part due to the guest OS memory manager's metadata that is explicitly allocated on each vNUMA node.
We find that the video workload sends fewer than 0.25\% of memory accesses to the \cvn node.
Similarly, the other three workloads send 0.06-0.38\% of memory access to the \cvn node.
Accesses within the local vNUMA node are spread out

\myparagraph{Implications}
With a negligible fraction of memory accesses on \cvn, we expect negligible performance impact given a correct prediction of untouched memory.

\subsection{\cvn VMs in the Lab}
\label{sec-eval-split}


\def \hmina {\hspace{-0.0in}}
\def \hminb {\hspace{-0.1in}}

\def \fgw {3.3in}
\def \fgh {1.085in}

\begin{figure}[t!]
\centerline{
\hmina
\includegraphics[width=\columnwidth]{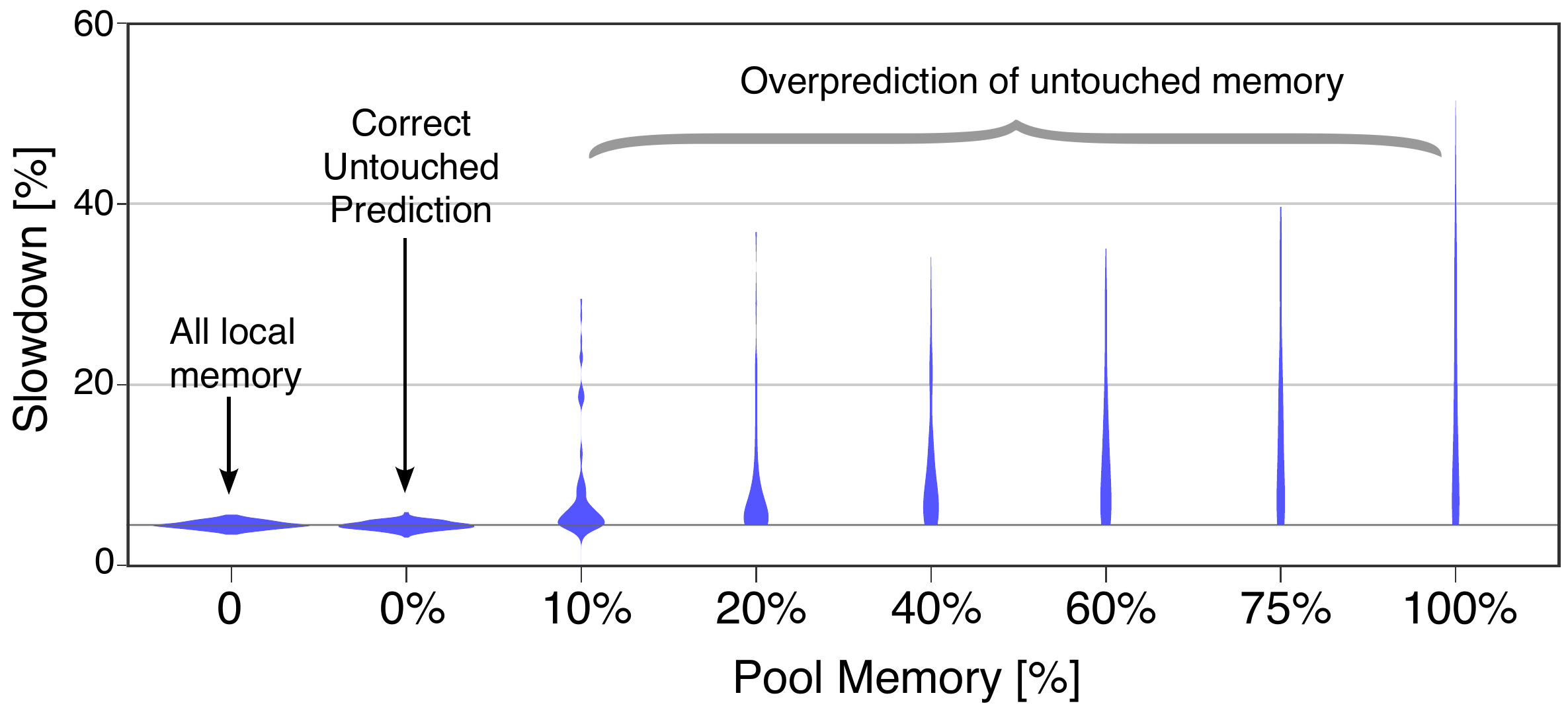}
}

%
    \mycaption{fig-split-cdf}{Slowdown under different pool allocations (\sec\ref{sec-eval-split})}{Performance for no pool memory and a correctly-sized zNUMA is comparable. Small slowdowns arise from run-to-run variation. Slowdown become noticable as soon as the workload spills into the zNUMA and steadily increases until the whole workload is allocated on pool memory (100\% spilled).}
\vten

\end{figure}

We scale up our evaluation to \numTotalApps\ workloads in a lab setting.
Since we fully control these workloads, we can now also explicitly measure their performance.
We rerun each workload on all-local memory, a correctly sized \cvn (0\% spilled), differently-sized \cvn nodes sized between 10-100\% of the workload's footprint.
Figure~\ref{fig-split-cdf} shows a violin plot of associated slowdowns.
This setup covers both normal behavior (all-local and 0\% spill) and misprediction behavior for latency sensitive workloads.
Thus, this is effectively a sensitivity study.

\myfinding{}
With a correct prediction of untouched memory, workload slowdowns have a similar distribution to all-local memory.

\myparagraph{Implications}
This performance result is expected since the \cvn node is rarely accessed (\sec\ref{sec:eval:accessbit}).
Our evaluation can thus assume no performance impact under correct predictions of untouched memory (\sec\ref{eval:e2e}).

\myfinding{}
For overpredictions of untouched memory (and correspondingly undersized local vNUMA nodes), the workload spills into \cvn.
Many workloads see an immediate impact on slowdown.
Slowdowns further increase if more workload memory spills into \cvn.
Some workloads are slowed down by as much as 30-35\% with 20-75\% of workload memory spilled and up to 50\% if entirely allocated on pool memory.
We use access bit scans to verify that these workloads indeed actively access their entire working set.

\myparagraph{Implications}
Allocating a fixed percentage of pool DRAM to VMs would lead to significant performance slowdowns.
There are only two strategies to reduce this impact: 1) identify which workloads will see slowdowns and 2) allocate untouched memory on the pool.
\sys employs both strategies.

\begin{figure}[t!]
\begin{center}
\includegraphics[width=0.95\columnwidth]{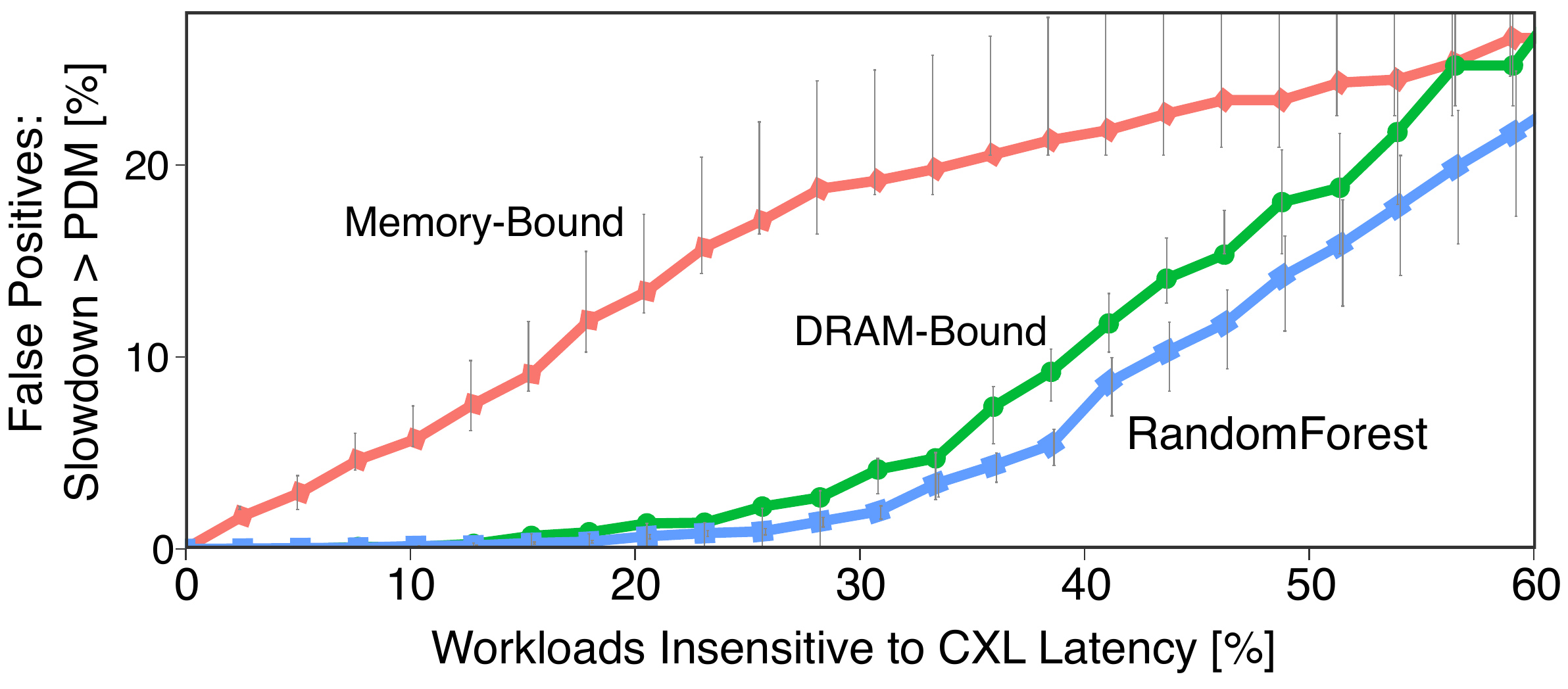}
    \mycaption{fig-ml-sensitivity}{Latency insensitivity model (\sec\ref{sec:eval:models})}{As we increase how many workloads are marked as insensitive (\texttt{LI}), the rate of false positives (\texttt{FP}) increases. \sys's RandomForest slightly outperforms a heuristic based only on the DRAM-bound TMA performance counter.}
    \vminfive
\end{center}
\end{figure}

\subsection{Performance of Prediction Models}
\label{sec:eval:models}

We evaluate \sys's prediction models (\sec\ref{sec:design:ml}) and its combined prediction model based on Eq.\eqref{eq:optimization}.

\vfive\subsubsection{Predicting Latency Sensitivity}
\label{sec-eval-pred}

\sys seeks to predict whether a VM is latency insensitive, \ie,
whether running the workload on pool memory would stay within the performance
degradation margin (\pdm).
We tested the model for \pdm between 1-10\% and on both 182\% and 222\% latency increases, but report details only for 5\% and 182\%.
Other \pdm values lead to qualitatively similar results.
\newtxt{The 222\% model is 16\% less effective given the same false positive rate target.}
We compare thresholds on
memory and DRAM boundedness~\cite{topdownanalysis.ispass14, tmam.web21} to
our RandomForest (\sec\ref{sec-impl}).

Figure~\ref{fig-ml-sensitivity}
shows the model's false positive rate as a function of the percentage of workloads
labeled as latency insensitive, similar to a precision-recall
curve~\cite{buckland1994relationship}.
Error bars show 99\% confidence from a 100-fold validation based on randomly
splitting into equal-sized training and testing datasets.

\myfinding{} While DRAM boundedness is correlated with
slowdown, we find examples where high slowdown occurs even for a small
percentage of DRAM boundedness. For example, multiple workloads exceed
20\% slowdown with just two percent of DRAM boundedness.

\myparagraph{Implication}
This shows the general hardness of predicting whether workloads exceed the \pdm.
Heuristic as well as predictors will make statistical errors.

\myfinding{}
We find that ``DRAM bound'' significantly outperforms ``Memory bound'' (Figure~\ref{fig-ml-sensitivity}).
Our RandomForest performs slightly better than ``DRAM bound''.

\myparagraph{Implication}
Our RandomForest can place 30\% of workloads on the pool with only 2\% of false positives.

\begin{figure}[t!]
\begin{center}
\includegraphics[width=0.95\columnwidth]{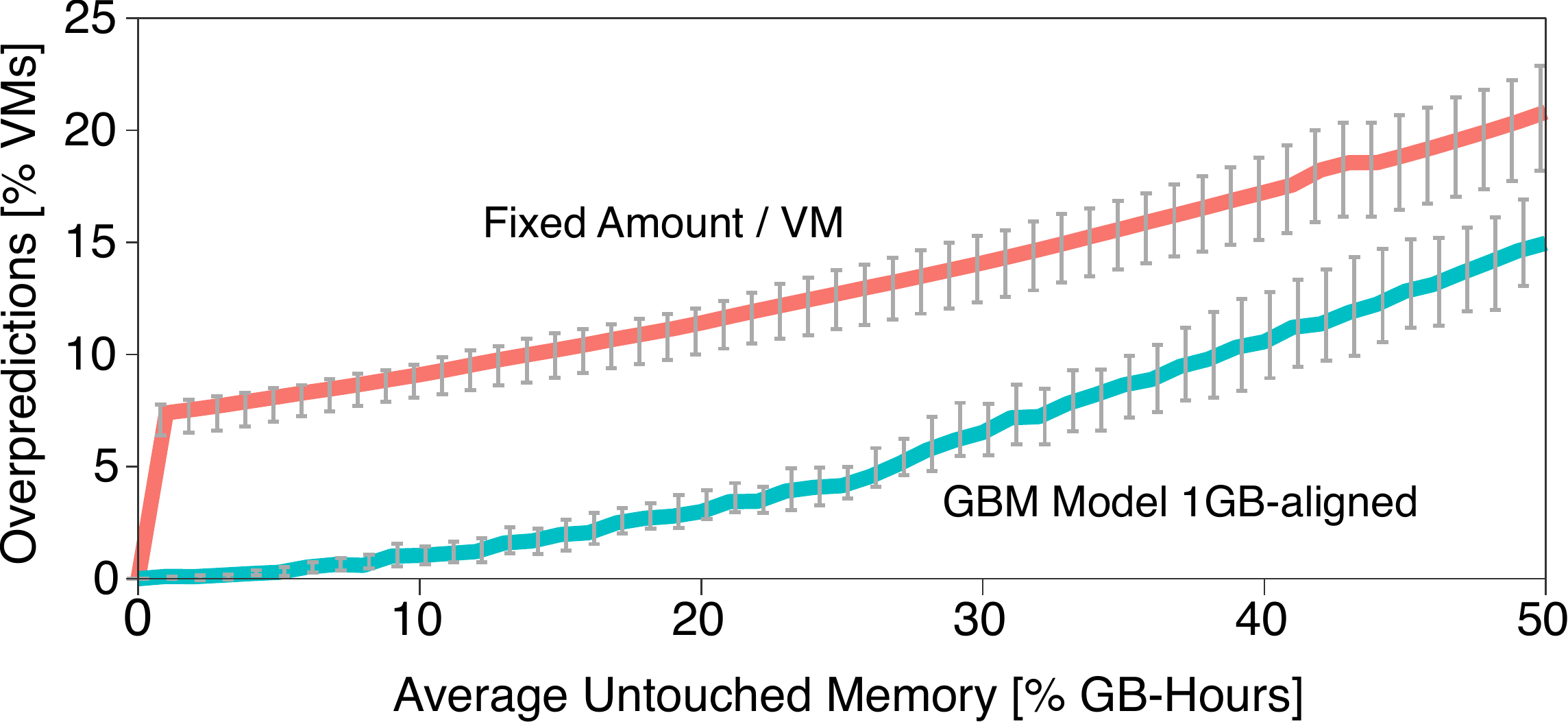}
    \mycaption{fig-ml-frigid}{Untouched memory model (\sec\ref{sec:eval:models})}{As we increase untouched memory (\texttt{UM}), our GBM has a significantly lower rate of overpredictions (\texttt{OP}) that a strawman model.}
    \vminfive
\end{center}
\end{figure}

\vfive\subsubsection{Predicting Untouched Memory}\label{eval:untouched}

\sys predicts the amount of untouched memory over a VM's future lifetime (\sec\ref{sec:design:ml}).
We evaluate this model using metadata and resource usage logs from 100 clusters over 75 days.
The model is trained nightly and evaluated on the subsequent day.
Figure~\ref{fig-ml-frigid} compares our GBM model to the heuristic that assumes a fixed fraction of memory as untouched across all VMs.
The figure shows the overprediction rate as a function of the average amount of untouched memory.
Figure~\ref{fig-ml-prod} shows a production version of the untouched memory model during the first 110 days of 2022.

\myfinding{}
\newtxt{We find that the GBM model is $5\times$ more accurate than the static policy, \eg, when labeling 20\% of memory as untouched,  GBM overpredicts only 2.5\% of VMs while the static policy overpredicts 12\%.}

\myparagraph{Implication}
\newtxt{Our prediction model identifies 25\% of untouched memory while only overpredicting 4\% of VMs.}

\myfinding{}
The production version of our model performs similarly to the simulated model.
Distributional shifts lead to some variability over time.

\myparagraph{Implication}
We find that accurately predicting untouched memory is practical and a realistic assumption.

\vfive\subsubsection{Combined Prediction Models}
\label{eval:combinedml}

\newtxt{We characterize \sys's combined models (Eq.\eqref{eq:optimization}) using ``scheduling mispredictions'', \ie, the fraction of VMs that will exceed the \pdm.
This incorporates the overpredictions of untouched memory, how much the model overpredicted, and the probability of this overprediction leading to a workload exceeding the \pdm.
Further, \sys uses its QoS monitor to mitigate up to 1\% of mispredictions.
Figure~\ref{fig-ml-combined} shows scheduling mispredictions as a function of the average amount of cluster DRAM that is allocated on its pools for 182\% and 222\% memory latency increases, respectively.}

\myfinding{}
\newtxt{\sys's combined model outperforms its individual models by finding their optimal combination.}
\myparagraph{Implication}
\newtxt{With a 2\%  scheduling misprediction target, \sys can schedule 44\% and 35\% of DRAM on pools with 182\% and 222\% memory latency increases, respectively.}


\begin{figure}[t]
\begin{center}
\includegraphics[width=\columnwidth]{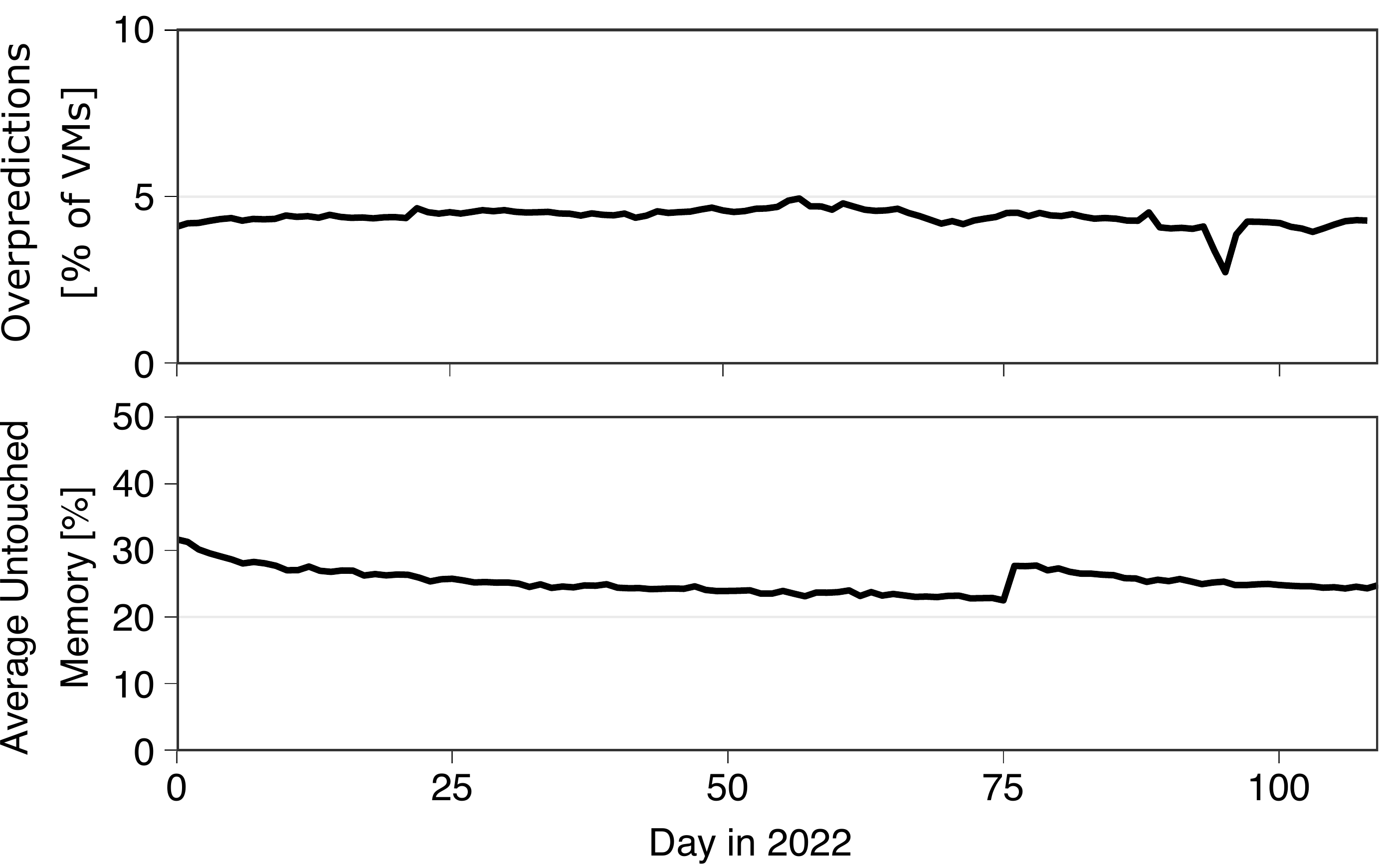} 
\vminten
    \mycaption{fig-ml-prod}{Untouched memory ML model performance in production (\sec\ref{sec:eval:models})}{\newtxt{Our production model targets 4\% overpredictions (\texttt{OP}). It average untouched memory percentage is similar to the simulated model (Figure~\ref{fig-ml-frigid}).}}
\vminten
\end{center}
\end{figure}

\subsection{End-to-end Reduction in Stranding}\label{eval:e2e}

\newtxt{We characterize \sys's end-to-end performance while constraining its rate of scheduling mispredictions.
Figure~\ref{fig-e2e} shows the reduction in aggregate cluster memory as a function of pool size for \sys under 182\% and 222\% memory latency increase, respectively, and a strawman static allocation policy.
We evaluate multiple scenarios; the figure shows \pdm=5\% and \tp=98\%.
In this scenario, the strawman statically allocates each VM with 15\% of pool DRAM.
About 10\% of VMs would touch the pool DRAM (Figure~\ref{fig-ml-frigid}).
Of those touching pool DRAM, we'd expect that about $\frac{1}{4}$ would see a slowdown exceeding a \pdm=5\% (Figure~\ref{fig-split-cdf}).
So, the strawman would have about 2.5\% of scheduling mispredictions.}

%

\myfinding{}
\newtxt{At a pool size of 16 sockets, \sys reduces overall DRAM requirements by 9\% and 7\% under 182\% and 222\% latency increases, respectively.
Static reduces DRAM by 3\%.
When varying \pdm between 1 and 10\% and \tp between 90 and 99.9\% we find the relative savings of the three systems to be qualitatively similar.}

\myparagraph{Implication}
\sys can safely reduce cost.
A QoS monitor that mitigates more than 1\% of mispredictions, can achieve more aggressive performance targets (\pdm).

\myfinding{}
Throughout the simulations, \sys's pool memory offlining speeds remain below 1GB/s and 10GB/s for 99.99\% and 99.999\% of VM starts, respectively.

\myparagraph{Implication}
\sys is practical and achieve design goals.

\begin{figure}[t]
\begin{center}
\vspace{-1mm}
\includegraphics[width=.8\columnwidth]{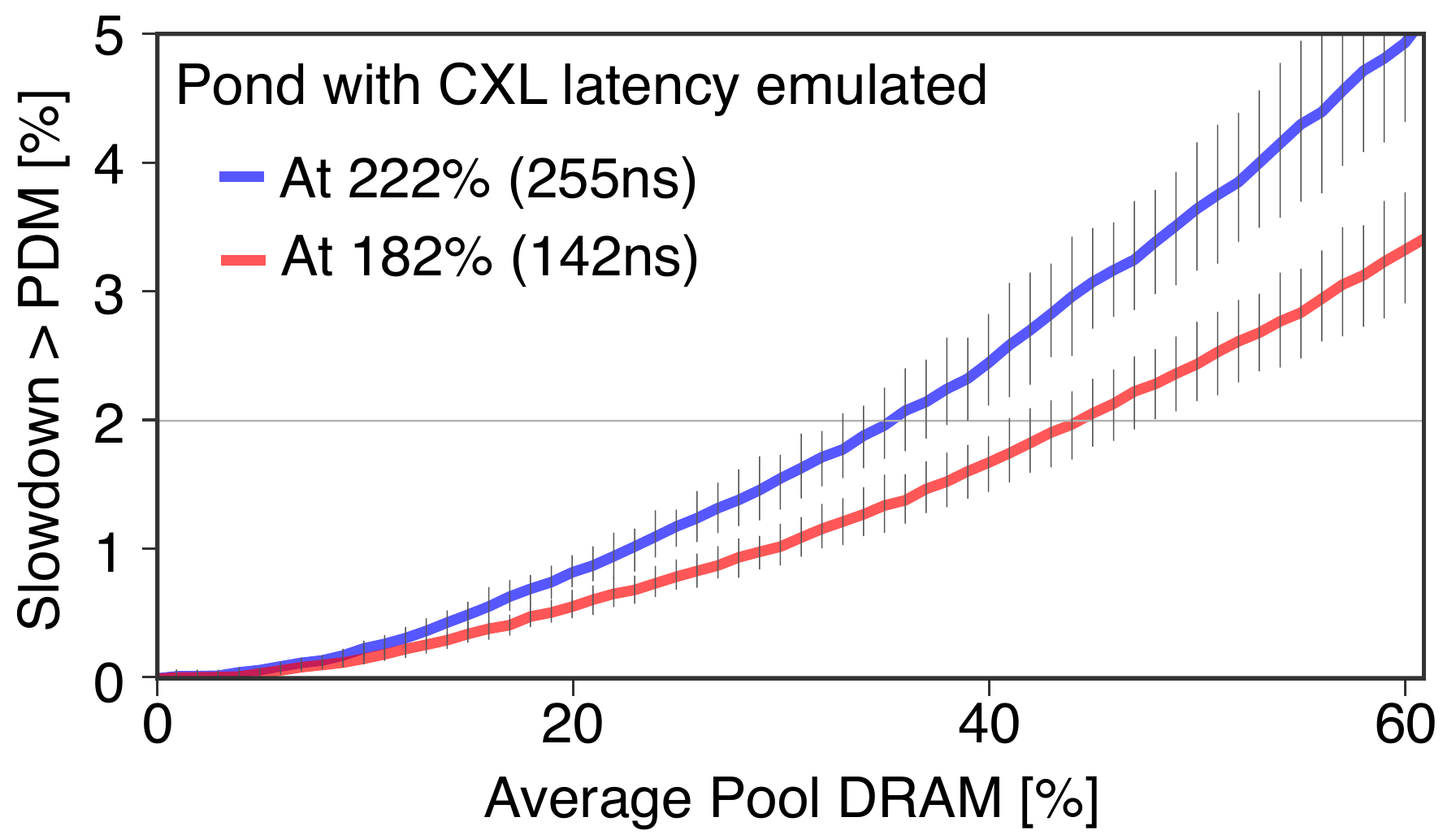}
    \mycaption{fig-ml-combined}{Combined model (\sec\ref{sec:eval:models})}{\newtxt{\sys's overall tradeoff between average allocation of pool memory and mispredictions after solving Eq.\eqref{eq:optimization}.}}
\end{center}
\vminten
\end{figure}

\section{Discussion}
\label{sec:discussion}

\myparagraph{Robustness of ML}
Similar to other oversubscribed resources (CPU~\cite{smartharvest.eurosys21} and disks~\cite{awsdisk}), customers may overuse resources to get local memory.
When multiplexing resource for millions of customers, any individual customer’s behavior will have a small impact.
Providers can also provide small discounts when resources are not fully utilized.

\myparagraph{Alternatives to static memory preallocation}
\sys is designed for compatibility static memory as potential workarounds are not yet practical.
The PCIe Address Translation Service (ATS/PRI)~\cite{pcieats.web09} enables compatibility with page faults.
Unfortunately, ATS/PRI-devices are not yet broadly available~\cite{nicpagefault.asplos17}.
Virtual IOMMUs~\cite{tian2020coiommu,amit2011viommu,ben2010turtles} allow fine-grained pinning but require guest OS changes and introduce overhead.

\section{Related Work}
\label{sec-rel}


{\ni\bf Hardware-level disaggregation:} Hardware-based
disaggregation designs~\cite{clio.web21, thymesisflow.micro20,
zombieland.eurosys18, pberry.hotos19, dredbox.date16, memblade.isca09,
optmemdisagg} are not easily deployable as they do not rely on commodity
hardware. For instance, ThymesisFlow~\cite{thymesisflow.micro20} and Clio~\cite{clio.web21}
propose FPGA-based rack-scale memory disaggregation designs on top of
{OpenCAPI}~\cite{opencapi.web21} and RDMA. Their hardware layer shares goals with \sys.
Their software goals differ fundamentally, \eg, ThymesisFlow advocates application changes
for performance, while \sys focuses on platform-level ML-driven pool
memory management that is transparent to users.



\vfive
{\ni\bf Hypervisor\slash OS level disaggregation:}
Hypervisor\slash OS level approaches~\cite{sysdisaggmem.hpca12,
    resdisagg.osdi16, infiniswap.nsdi17, legoos.osdi18,
    softfarmem.asplos19, dcm.tc19, fastswap.eurosys20, leap.atc20,
fluidmem.icdcs20, orchdisaggmem.tc20, ememdisagg.corr20}
rely on page faults and access monitoring to maintain the
working set in local DRAM. Such OS-based approaches bring significant
overhead, jitter, and are incompatible with virtualization acceleration
(\eg, DDA).

%


\begin{figure}[t]
\begin{center}
\vspace{-1mm}
\includegraphics[width=.9\columnwidth]{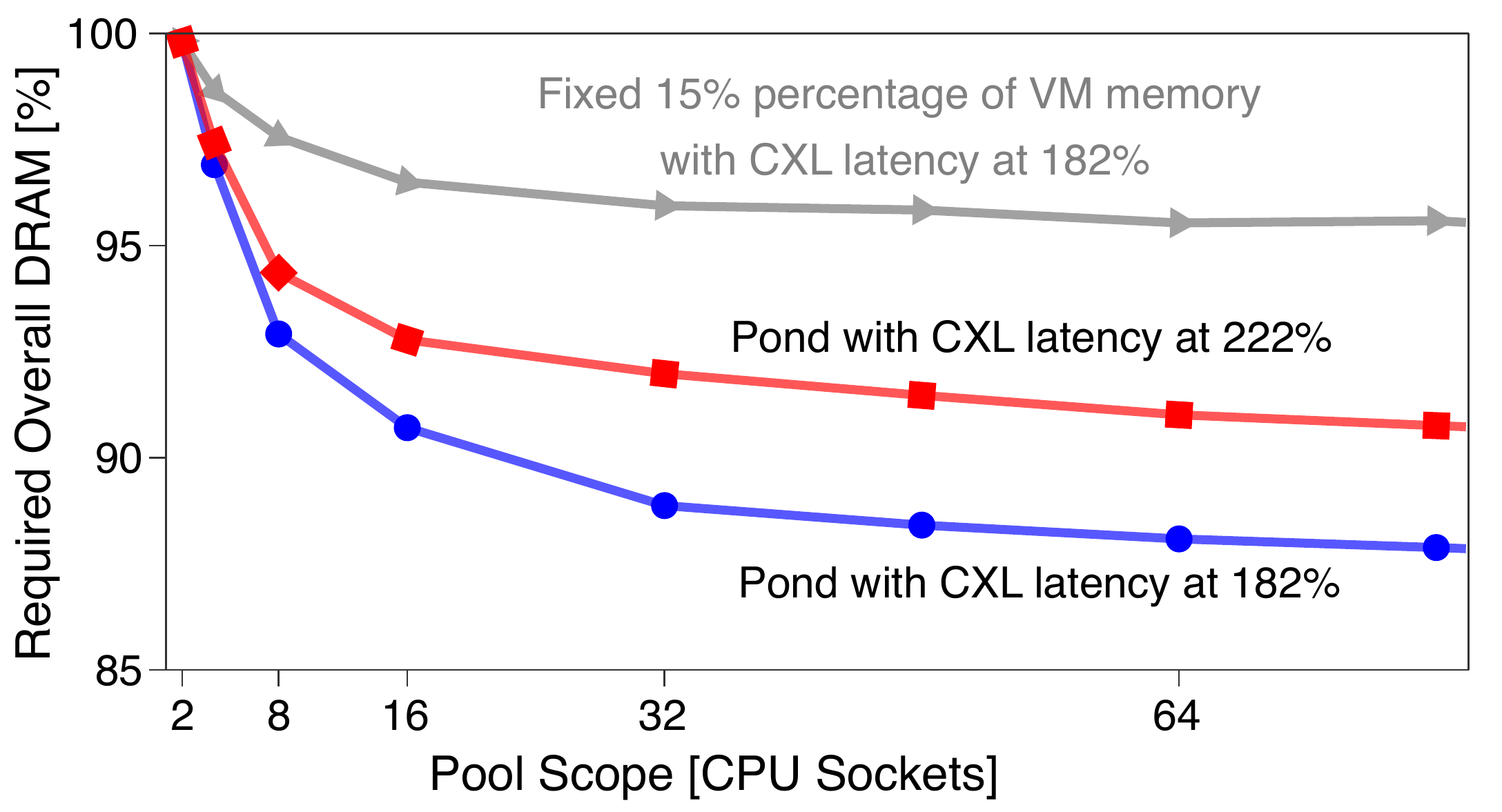}
\mycaption{fig-e2e}{Memory savings under performance constraints (\sec\ref{eval:e2e})}
{\newtxt{Simulated end-to-end evaluation of memory savings achieved by \sys under \pdm=5\% and scheduling mispredictions \tp=98\%.}}
\vminten
\end{center}
\end{figure}

\vfive
{\ni\bf Runtime/application level disaggregation:}
Runtime-based disaggregation designs~\cite{aifm.osdi20, semeru.osdi20,
kona.asplos21} propose customized APIs for remote memory access.
While effective, this approach requires
developers to explicitly use these mechanisms at the application level.

\vfive
{\ni\bf Memory tiering:}
Prior works have considered the broader impact of extended memory
hierarchies and how to handle them~\cite{softfarmem.asplos19,
heteroos.isca17, thermostat.asplos17, flatflash.asplos19,
autotiering.atc21}. For example,
Google
achieves 6\us\ latency via proactive hot\slash cold page
detection and compression~\cite{zswap.web20, softfarmem.asplos19}.
%
Nimble~\cite{nimblepage.asplos19} optimizes Linux's page tracking mechanism to
tier pages for increased migration bandwidth.
%
%
\sys takes a different ML-based approach looking at memory pooling design at the platform-level and is orthogonal
to these works.
%

\vfive
{\ni\bf ML for systems:}
ML is increasingly applied to tackle systems problems,
such as cloud efficiency~\cite{smartharvest.eurosys21,
resourcecentral.sosp17}, memory\slash storage
optimizations~\cite{mldistore.mlsys21, llama.asplos20},
microservices~\cite{sinan.asplos21}, caching\slash prefetching policies
~\cite{nmprefetch.asplos21, dlcache.micro19}. We uniquely apply ML methods
for untouched memory prediction to support pooled memory provisioning to VMs
without jeopardizing QoS.

\vfive
{\ni\bf Coherent memory and NUMA optimizations:}
Traditional cache coherent NUMA architectures~\cite{ccnuma.isca97} use
specialized interconnects to implement a shared address space.
There are also system-level optimizations for NUMA, such as
NUMA-aware data placement~\cite{asymmemplacement.atc15} and proactive page
migration~\cite{autonuma.web19}.
%
NUMA scheduling policies ~\cite{corbet2012autonuma,rao2010vnuma,liu2014optimizing} balance compute and memory across NUMA nodes.
\sys's ownership overcomes the need for coherence across the memory pool.
\cvn's zero-core nature
requires rethinking of existing optimizations which are largely
optimized for symmetric NUMA systems.
\section{Conclusion}

DRAM costs are an increasing cost factor for cloud providers.
This paper is motivated by the observation of stranded and untouched memory across 100 production cloud clusters.
We proposed \sys, the first full-stack memory pool that satisfies the requirements of cloud providers.
\sys
comprises contributions at the hardware, system software, and
distributed system layers.
Our results showed
that \sys can reduce the amount of needed DRAM by 7\% with a pool size of 16 sockets and assuming CXL increases latency by 222\%.
This translates into an overall reduction of 3.5\% in cloud server cost.



\newpage
\begingroup
\setcounter{page}{1}


\singlespacing
\setlength{\bibsep}{1pt plus 0.1ex}
{\footnotesize

   \bibliographystyle{unsrt} 


   \bibliography{B/defs,B/all,B/personal,local.bib,mem.bib,B/confs}

\begin{thebibliography}{100}

\bibitem{nicpagefault.asplos17}
Ilya Lesokhin, Haggai Eran, Shachar Raindel, Guy Shapiro, Sagi Grimberg, Liran
  Liss, Muli Ben-Yehuda, Nadav Amit, and Dan Tsafrir.
\newblock {Page Fault Support for Network Controllers}.
\newblock In {\em Proceedings of the 22nd ACM International Conference on
  Architectural Support for Programming Languages and Operating Systems
  (ASPLOS)}, 2017.

\bibitem{tian2020coiommu}
Kun Tian, Yu~Zhang, Luwei Kang, Yan Zhao, and Yaozu Dong.
\newblock {coIOMMU: A Virtual IOMMU with Cooperative DMA Buffer Tracking for
  Efficient Memory Management in Direct I/O}.
\newblock In {\em Proceedings of the 2020 USENIX Annual Technical Conference
  (ATC)}, 2020.

\bibitem{yassour2010dma}
Ben-Ami Yassour, Muli Ben-Yehuda, and Orit Wasserman.
\newblock {On the DMA Mapping Problem in Direct Device Assignment}.
\newblock In {\em Proceedings of the 3rd ACM International Conference on
  Systems and Storage (SYSTOR)}, 2010.

\bibitem{willmann2008protection}
Paul Willmann, Scott Rixner, and Alan~L. Cox.
\newblock {Protection Strategies for Direct Access to Virtualized I/O Devices}.
\newblock In {\em Proceedings of the USENIX Annual Technical Conference (ATC)},
  2008.

\bibitem{amit2011viommu}
Nadav Amit, Muli Ben-Yehuda, IBM Research, Dan Tsafrir, and Assaf Schuster.
\newblock {vIOMMU: Efficient IOMMU Emulation}.
\newblock In {\em Proceedings of the 2011 USENIX Annual Technical Conference
  (ATC)}, 2011.

\bibitem{ben2010turtles}
Muli Ben-Yehuda, Michael~D. Day, Zvi Dubitzky, Michael Factor, Nadav Har’El,
  Abel Gordon, Anthony Liguori, Orit Wasserman, and Ben-Ami Yassour.
\newblock {The Turtles Project: Design and Implementation of Nested
  Virtualization}.
\newblock In {\em Proceedings of the 9th USENIX Symposium on Operating Systems
  Design and Implementation (OSDI)}, 2010.

\bibitem{awsaccelnet}
{AWS: Enhanced networking support}.
\newblock
  \url{https://docs.aws.amazon.com/AWSEC2/latest/UserGuide/enhanced-networking.html#supported_instances}.

\bibitem{azureaccelnet}
{Azure Accelerated Networking: Supported VM instances}.
\newblock
  \url{https://docs.microsoft.com/en-us/azure/virtual-network/accelerated-networking-overview#supported-vm-instances}.

\bibitem{memscaling.imw13}
Onur Mutlu.
\newblock {Memory Scaling: A Systems Architecture Perspective}.
\newblock In {\em IEEE International Memory Workshop (IMW)}, 2013.

\bibitem{archshield.isca13}
Prashant Nair, Dae-Hyun Kim, and Moinuddin~K. Qureshi.
\newblock {ArchShield: Architectural Framework for Assisting DRAM Scaling by
  Tolerating High Error Rates}.
\newblock In {\em Proceedings of the 40th Annual International Symposium on
  Computer Architecture (ISCA)}, 2013.

\bibitem{mutlu2015main}
Onur Mutlu.
\newblock {Main Memory Scaling: Challenges and Solution directions}.
\newblock In {\em {More than Moore Technologies for Next Generation Computer
  Design}}. Springer, 2015.

\bibitem{dramscaling.imw15}
Sung-Kye Park.
\newblock {Technology Scaling Challenge and Future Prospects of DRAM and NAND
  Flash Memory}.
\newblock In {\em IEEE International Memory Workshop (IMW)}, 2015.

\bibitem{dramscalingchallenges.imw20}
Shigeru Shiratake.
\newblock {Scaling and Performance Challenges of Future DRAM}.
\newblock In {\em IEEE International Memory Workshop (IMW)}, 2020.

\bibitem{nextnewmemories.web19}
{The Next New Memories}.
\newblock \url{https://semiengineering.com/the-next-new-memories/}, 2019.

\bibitem{micron3dxp.news21}
{Micron Ends 3D XPoint Memory}.
\newblock
  \url{https://www.forbes.com/sites/tomcoughlin/2021/03/16/micron-ends-3d-xpoint-memory/},
  2021.

\bibitem{cxlandgenz.web20}
{CXL And Gen-Z Iron Out A Coherent Interconnect Strategy}.
\newblock
  \url{https://www.nextplatform.com/2020/04/03/cxl-and-gen-z-iron-out-a-coherent-interconnect-strategy/},
  2020.

\bibitem{softfarmem.asplos19}
Andres Lagar-Cavilla, Junwhan Ahn, Suleiman Souhlal, Neha Agarwal, Radoslaw
  Burny, Shakeel Butt, Jichuan Chang, Ashwin Chaugule, Nan Deng, Junaid Shahid,
  Greg Thelen, Kamil~Adam Yurtsever, Yu~Zhao, and Parthasarathy Ranganathan.
\newblock {Software-Defined Far Memory in Warehouse-Scale Computers}.
\newblock In {\em Proceedings of the 24th ACM International Conference on
  Architectural Support for Programming Languages and Operating Systems
  (ASPLOS)}, 2019.

\bibitem{weiner2022tmo}
Johannes Weiner, Niket Agarwal, Dan Schatzberg, Leon Yang, Hao Wang, Blaise
  Sanouillet, Bikash Sharma, Tejun Heo, Mayank Jain, Chunqiang Tang, and
  Dimitrios Skarlatos.
\newblock {TMO: Transparent Memory Offloading in Datacenters}.
\newblock In {\em Proceedings of the 27th ACM International Conference on
  Architectural Support for Programming Languages and Operating Systems
  (ASPLOS)}, 2022.

\bibitem{memblade.isca09}
Kevin~T. Lim, Jichuan Chang, Trevor~N. Mudge, Parthasarathy Ranganathan,
  Steven~K. Reinhardt, and Thomas~F. Wenisch.
\newblock {Disaggregated Memory for Expansion and Sharing in Blade Servers}.
\newblock In {\em Proceedings of the 36th Annual International Symposium on
  Computer Architecture (ISCA)}, 2009.

\bibitem{lim2011disaggregated}
Kevin Lim, Yoshio Turner, Jichuan Chang, J~Renato Santos, and Parthasarathy
  Ranganathan.
\newblock {Disaggregated Memory Benefits for Server Consolidation}.
\newblock {\em HP Laboratories}, 2011.

\bibitem{orchdisaggmem.tc20}
Wenqi Cao and Ling Liu.
\newblock {Hierarchical Orchestration of Disaggregated Memory}.
\newblock {\em IEEE Transactions on Computers (TC)}, 69(6), June 2020.

\bibitem{sysdisaggmem.hpca12}
Kevin Lim, Yoshio Turner, Jose~Renato Santos, Alvin AuYoung, Jichuan Chang,
  Parthasarathy Ranganathan, and Thomas~F. Wenisch.
\newblock {System-level Implications of Disaggregated Memory}.
\newblock In {\em Proceedings of the 18th International Symposium on High
  Performance Computer Architecture (HPCA-18)}, 2012.

\bibitem{aifm.osdi20}
Zhenyuan Ruan, Malte Schwarzkopf, Marcos~K. Aguilera, and Adam Belay.
\newblock {AIFM: High-Performance, Application-Integrated Far Memory}.
\newblock In {\em Proceedings of the 14th USENIX Symposium on Operating Systems
  Design and Implementation (OSDI)}, 2020.

\bibitem{angel2020disaggregation}
Sebastian Angel, Mihir Nanavati, and Siddhartha Sen.
\newblock {Disaggregation and the Application}.
\newblock In {\em The 12th USENIX Workshop on Hot Topics in Cloud Computing
  (HotCloud)}, 2020.

\bibitem{semeru.osdi20}
Chenxi Wang, Haoran Ma, Shi Liu, Yuanqi Li, Zhenyuan Ruan, Khanh Nguyen,
  Michael~D. Bond, Ravi Netravali, Miryung Kim, and Guoqing~Harry Xu.
\newblock {Semeru: A Memory-Disaggregated Managed Runtime}.
\newblock In {\em Proceedings of the 14th USENIX Symposium on Operating Systems
  Design and Implementation (OSDI)}, 2020.

\bibitem{kona.asplos21}
Irina Calciu, M.~Talha Imran, Ivan Puddu, Sanidhya Kashyap, Hasan~Al Maruf,
  Onur Mutlu, and Aasheesh Kolli.
\newblock {Rethinking Software Runtimes for Disaggregated Memory}.
\newblock In {\em Proceedings of the 26th ACM International Conference on
  Architectural Support for Programming Languages and Operating Systems
  (ASPLOS)}, 2021.

\bibitem{farm.nsdi14}
Aleksandar Dragojevic, Dushyanth Narayanan, Miguel Castro, and Orion Hodson.
\newblock {FaRM: Fast Remote Memory}.
\newblock In {\em Proceedings of the 11th USENIX Symposium on Networked Systems
  Design and Implementation (NSDI)}, 2014.

\bibitem{remoteregions.atc18}
Marcos~K. Aguilera, Nadav Amit, Irina Calciu, Xavier Deguillard, Jayneel
  Gandhi, Stanko Novakovic, Arun Ramanathan, Pratap Subrahmanyam, Lalith
  Suresh, Kiran Tati, Rajesh Venkatasubramanian, and Michael Wei.
\newblock {Remote Regions: a Simple Abstraction for Remote Memory}.
\newblock In {\em Proceedings of the 2018 USENIX Annual Technical Conference
  (ATC)}, 2018.

\bibitem{kvdisagg.atc20}
Shin-Yeh Tsai, Yizhou Shan, and Yiying Zhang.
\newblock {Disaggregating Persistent Memory and Controlling Them Remotely: An
  Exploration of Passive Disaggregated Key-Value Stores}.
\newblock In {\em Proceedings of the 2020 USENIX Annual Technical Conference
  (ATC)}, 2020.

\bibitem{farview.corr21}
Dario Korolija, Dimitrios Koutsoukos, Kimberly Keeton, Konstantin Taranov,
  Dejan Milojicic, and Gustavo Alonso.
\newblock {Farview: Disaggregated Memory with Operator Offloading for Database
  Engines}.
\newblock {\em arXiv:2106.07102}, 2021.

\bibitem{resdisagg.osdi16}
Peter~X. Gao, Akshay Narayan, Sagar Karandikar, Joao Carreira, Sangjin Han,
  Rachit Agarwal, Sylvia Ratnasamy, and Scott Shenker.
\newblock {Network Requirements for Resource Disaggregation}.
\newblock In {\em Proceedings of the 12th USENIX Symposium on Operating Systems
  Design and Implementation (OSDI)}, 2016.

\bibitem{infiniswap.nsdi17}
Juncheng Gu, Youngmoon Lee, Yiwen Zhang, Mosharaf Chowdhury, and Kang~G. Shin.
\newblock {Efficient Memory Disaggregation with Infiniswap}.
\newblock In {\em Proceedings of the 14th USENIX Symposium on Networked Systems
  Design and Implementation (NSDI)}, 2017.

\bibitem{legoos.osdi18}
Yizhou Shan, Yutong Huang, Yilun Chen, and Yiying Zhang.
\newblock {LegoOS: A Disseminated, Distributed OS for Hardware Resource
  Disaggregation}.
\newblock In {\em Proceedings of the 13th USENIX Symposium on Operating Systems
  Design and Implementation (OSDI)}, 2018.

\bibitem{dcm.tc19}
Kwangwon Koh, Kangho Kim, Seunghyub Jeon, and Jaehyuk Huh.
\newblock {Disaggregated Cloud Memory with Elastic Block Management}.
\newblock {\em IEEE Transactions on Computers (TC)}, 68(1), 2019.

\bibitem{fastswap.eurosys20}
Emmanuel Amaro, Christopher Branner-Augmon, Zhihong Luo, Amy Ousterhout,
  Marcos~K. Aguilera, Aurojit Panda, Sylvia Ratnasamy, and Scott Shenker.
\newblock {Can Far Memory Improve Job Throughput?}
\newblock In {\em Proceedings of the 2020 EuroSys Conference (EuroSys)}, 2020.

\bibitem{leap.atc20}
Hasan~Al Maruf and Mosharaf Chowdhury.
\newblock {Effectively Prefetching Remote Memory with Leap}.
\newblock In {\em Proceedings of the 2020 USENIX Annual Technical Conference
  (ATC)}, 2020.

\bibitem{fluidmem.icdcs20}
Blake Caldwell, Sepideh Goodarzy, Sangtae Ha, Richard Han, Eric Keller, Eric
  Rozner, and Youngbin Im.
\newblock {FluidMem: Full, Flexible, and Fast Memory Disaggregation for the
  Cloud}.
\newblock In {\em International Conference on Distributed Computing Systems
  (ICDCS)}, 2020.

\bibitem{ememdisagg.corr20}
Youngmoon Lee, Hassan~Al Maruf, Mosharaf Chowdhury, and Kang~G. Shin.
\newblock {Mitigating the Performance-Efficiency Tradeoff in Resilient Memory
  Disaggregation}.
\newblock {\em arXiv:1910.09727}, 2019.

\bibitem{cxlsite.web20}
{Compute Express Link Specification}.
\newblock Available at \url{https://www.computeexpresslink.org}, 2020.

\bibitem{intelsapphire.web21}
{Sapphire Rapids Uncovered: 56 Cores, 64GB HBM2E, Multi-Chip Design}.
\newblock
  \url{https://www.tomshardware.com/news/intel-sapphire-rapids-xeon-scalable-specifications-and-features},
  2021.

\bibitem{arm2021cxl}
{CXL Consortium Member Spotlight: Arm}.
\newblock
  \url{https://www.computeexpresslink.org/post/cxl-consortium-member-spotlight-arm},
  2021.

\bibitem{amdgenoa}
{AMD Unveils Workload-Tailored Innovations and Products at The Accelerated Data
  Center Premiere}.
\newblock
  \url{https://www.amd.com/en/press-releases/2021-11-08-amd-unveils-workload-tailored-innovations-and-products-the-accelerated},
  2021.

\bibitem{cpulessnuma.web19}
Christopher Lameter.
\newblock {Flavors of Memory supported by Linux, Their Use and Benefit}.
\newblock
  \url{https://events19.linuxfoundation.org/wp-content/uploads/2017/11/The-Flavors-of-Memory-Supported-by-Linux-their-Use-and-Benefit-Christoph-Lameter-Jump-Trading-LLC.pdf},
  2019.

\bibitem{firecracker.nsdi20}
Alexandru Agache, Marc Brooker, Andreea Florescu, Alexandra Iordache, Anthony
  Liguori, Rolf Neugebauer, Phil Piwonka, and Diana-Maria Popa.
\newblock {Firecracker: Lightweight Virtualization for Serverless
  Applications}.
\newblock In {\em Proceedings of the 17th USENIX Symposium on Networked Systems
  Design and Implementation (NSDI)}, 2020.

\bibitem{intelvtd.web20}
{Intel Virtualization Technology for Directed I/O}.
\newblock
  \url{https://software.intel.com/content/dam/develop/external/us/en/documents/vt-directed-io-spec.pdf},
  2020.

\bibitem{leapio.asplos20}
Huaicheng Li, Mingzhe Hao, Stanko Novakovic, Vaibhav Gogte, Sriram Govindan,
  Dan R.~K. Ports, Irene Zhang, Ricardo Bianchini, Haryadi~S. Gunawi, and
  Anirudh Badam.
\newblock {LeapIO: Efficient and Portable Virtual NVMe Storage on ARM SoCs}.
\newblock In {\em Proceedings of the 25th ACM International Conference on
  Architectural Support for Programming Languages and Operating Systems
  (ASPLOS)}, 2020.

\bibitem{sriov}
{Single-Root Input/Output Virtualization}.
\newblock \url{httpp://www.pcisig.com/specifications/iov/single_root}, 2019.

\bibitem{resourcecentral.sosp17}
Eli Cortez, Anand Bonde, Alexandre Muzio, Mark Russinovich, Marcus Fontoura,
  and Ricardo Bianchini.
\newblock {Resource Central: Understanding and Predicting Workloads for
  Improved Resource Management in Large Cloud Platforms}.
\newblock In {\em Proceedings of the 26th ACM Symposium on Operating Systems
  Principles (SOSP)}, 2017.

\bibitem{hadary2020protean}
Ori Hadary, Luke Marshall, Ishai Menache, Abhisek Pan, Esaias~E Greeff, David
  Dion, Star Dorminey, Shailesh Joshi, Yang Chen, Mark Russinovich, and Thomas
  Moscibroda.
\newblock {Protean: VM Allocation Service at Scale}.
\newblock In {\em Proceedings of the 14th USENIX Symposium on Operating Systems
  Design and Implementation (OSDI)}, 2020.

\bibitem{googlejobpacking.cluster14}
Abhishek Verma, Madhukar Korupolu, and John Wilkes.
\newblock {Evaluating Job Packing in Warehouse-scale Computing}.
\newblock In {\em International Conference on Cluster Computing (Cluster)},
  2014.

\bibitem{borg.eurosys15}
Abhishek Verma, Luis Pedrosa, Madhukar Korupolu, David Oppenheimer, Eric Tune,
  and John Wilkes.
\newblock {Large-scale cluster management at Google with Borg}.
\newblock In {\em Proceedings of the 2015 EuroSys Conference (EuroSys)}, 2015.

\bibitem{binpackheuristics.web11}
Rina Panigrahy, Kunal Talwar, Lincoln Uyeda, and Udi Wieder.
\newblock {Heuristics for Vector Bin Packing}.
\newblock
  \url{https://www.microsoft.com/en-us/research/publication/heuristics-for-vector-bin-packing/},
  2011.

\bibitem{tetris.sigcomm14}
Robert Grandl, Ganesh Ananthanarayanan, Srikanth Kandula, Sriram Rao, and
  Aditya Akella.
\newblock {Multi-Resource Packing for Cluster Schedulers}.
\newblock In {\em Proceedings of the ACM Special Interest Group on Data
  Communication (SIGCOMM)}, 2014.

\bibitem{bpbounds.stoc13}
Yossi Azar, Ilan~Reuven Cohen, Seny Kamara, and Bruce Shepherd.
\newblock {Tight Bounds for Online Vector Bin Packing}.
\newblock In {\em Proceedings of the 45th ACM symposium on Theory of Computing
  (STOC)}, 2013.

\bibitem{smartharvest.eurosys21}
Yawen Wang, Kapil Arya, Marios Kogias, Manohar Vanga, Aditya Bhandari,
  Neeraja~J. Yadwadkar, Siddhartha Sen, Sameh Elnikety, Christos Kozyrakis, and
  Ricardo Bianchini.
\newblock {SmartHarvest: Harvesting Idle CPUs Safely and Efficiently in the
  Cloud}.
\newblock In {\em Proceedings of the 2021 EuroSys Conference (EuroSys)}, 2021.

\bibitem{harvestslo.osdi20}
Pradeep Ambati, Íñigo Goiri, Felipe~Vieira Frujeri, Alper Gun, Ke~Wang, Brian
  Dolan, Brian Corell, Sekhar Pasupuleti, Thomas Moscibroda, Sameh Elnikety,
  Marcus Fontoura, and Ricardo Bianchini.
\newblock {Providing SLOs for Resource-Harvesting VMs in Cloud Platforms}.
\newblock In {\em Proceedings of the 14th USENIX Symposium on Operating Systems
  Design and Implementation (OSDI)}, 2020.

\bibitem{fastnetdisagg.socc17}
Marcos~K. Aguilera, Nadav Amit, Irina Calciu, Xavier Deguillard, Jayneel
  Gandhi, Pratap Subrahmanyam, Lalith Suresh, Kiran Tati, Rajesh
  Venkatasubramanian, and Michael Wei.
\newblock {Remote Memory in the Age of Fast Networks}.
\newblock In {\em Proceedings of the 8th ACM Symposium on Cloud Computing
  (SoCC)}, 2017.

\bibitem{thymesisflow.micro20}
Christian Pinto, Dimitris Syrivelis, Michele Gazzetti, Panos Koutsovasilis,
  Andrea Reale, Kostas Katrinis, and Peter Hofstee.
\newblock {ThymesisFlow: A Software-Defined, HW/SW Co-Designed Interconnect
  Stack for Rack-Scale Memory Disaggregation}.
\newblock In {\em 53rd Annual IEEE/ACM International Symposium on
  Microarchitecture (MICRO-53)}, 2020.

\bibitem{cxl2spec.web20}
{CXL 2.0 Specification}.
\newblock \url{https://www.computeexpresslink.org/download-the-specification},
  2020.

\bibitem{cxl2whitepaper.web21}
{Compute Express Link 2.0 White Paper}.
\newblock
  \url{https://b373eaf2-67af-4a29-b28c-3aae9e644f30.filesusr.com/ugd/0c1418_14c5283e7f3e40f9b2955c7d0f60bebe.pdf},
  2021.

\bibitem{debendra2022fms}
Debendra~Das Sharma.
\newblock {CXL 3.0: New Features for Increased Scale and Optimized Resource
  Utilization}.
\newblock In {\em Flash Memory Summit}, 2022.

\bibitem{cxl3spec}
{CXL 3.0 Specification}.
\newblock \url{https://www.computeexpresslink.org/download-the-specification},
  2022.

\bibitem{debendra2022hoti}
Debendra~Das Sharma.
\newblock {{Compute Express Link}: An Open Industry-standard Interconnect
  Enabling Heterogenous Data-centric Computing}.
\newblock In {\em Proceedings of the 29th IEEE Hot Interconnects symposium
  (HotI29)}, 2022.

\bibitem{maruf2022tpp}
Hasan~Al Maruf, Hao Wang, Abhishek Dhanotia, Johannes Weiner, Niket Agarwal,
  Pallab Bhattacharya, Chris Petersen, Mosharaf Chowdhury, Shobhit Kanaujia,
  and Prakash Chauhan.
\newblock {TPP: Transparent Page Placement for CXL-Enabled Tiered Memory}.
\newblock {\em arXiv:2206.02878}, 2022.

\bibitem{gouk2022direct}
Donghyun Gouk, Sangwon Lee, Miryeong Kwon, , and Myoungsoo Jung.
\newblock {Direct Access, High-Performance Memory Disaggregation with
  DirectCXL}.
\newblock In {\em Proceedings of the 2022 USENIX Annual Technical Conference
  (ATC)}, 2022.

\bibitem{genoa2021leak}
{AMD EPYC Genoa and SP5 Platform Leaked}.
\newblock
  \url{https://docs.aws.amazon.com/AWSEC2/latest/UserGuide/disk-performance.html},
  2021.

\bibitem{intelras}
{Reliability, Availability, and Serviceability (RAS) Integration and Validation
  Guide for the Intel Xeon Processor E7 Family}.
\newblock
  \url{https://www.intel.com/content/dam/develop/external/us/en/documents/emca2-integration-validation-guide-556978.pdf},
  2015.

\bibitem{amdras}
{AMD EPYC brings new RAS capability}.
\newblock
  \url{https://www.amd.com/system/files/2017-06/AMD-EPYC-Brings-New-RAS-Capability.pdf},
  2017.

\bibitem{microchipretimer}
{CXL Use-cases Driving the Need For Low Latency Performance Retimers}.
\newblock
  \url{https://www.microchip.com/en-us/about/blog/learning-center/cxl--use-cases-driving-the-need-for-low-latency-performance-reti},
  2021.

\bibitem{asteraretimer}
{Enabling PCIe 5.0 System Level Testing and Low Latency Mode for CXL}.
\newblock
  \url{https://www.asteralabs.com/videos/aries-smart-retimer-for-pcie-gen-5-and-cxl/},
  2021.

\bibitem{cxlretimers}
Elene Chobanyan, Casey Morrison, and Pegah Alavi.
\newblock {End-to-End System-Level Simulations with Retimers for PCIe Gen5 \&
  CXL}.
\newblock DesignCon, slides available at
  \url{https://www.asteralabs.com/wp-content/themes/astera-labs/images/retimer-cxl.pdf},
  2020.

\bibitem{cxlswitchlatency}
Timothy~Prickett Morgan.
\newblock Pci-express 5.0: The unintended but formidable datacenter
  interconnect.
\newblock DesignCon, slides available at
  \url{https://www.nextplatform.com/2021/02/03/pci-express-5-0-the-unintended-but-formidable-datacenter-interconnect/},
  2021.

\bibitem{i3c}
{MIPI I3C Bus Sensor Specification}.
\newblock \url{https://www.mipi.org/specifications/i3c-sensor-specification},
  2021.

\bibitem{acpi}
UEFI.
\newblock {Advanced Configuration and Power Interface Specification}, 2021.

\bibitem{nimblepage.asplos19}
Zi~Yan, Daniel Lustig, David Nellans, and Abhishek Bhattacharjee.
\newblock {Nimble Page Management for Tiered Memory Systems}.
\newblock In {\em Proceedings of the 24th ACM International Conference on
  Architectural Support for Programming Languages and Operating Systems
  (ASPLOS)}, 2019.

\bibitem{autonuma.web19}
Huang Ying.
\newblock {AutoNUMA: Optimize Memory Placement for Memory Tiering System}.
\newblock \url{https://lwn.net/Articles/835402/}, 2019.

\bibitem{ruprecht2018vm}
Adam Ruprecht, Danny Jones, Dmitry Shiraev, Greg Harmon, Maya Spivak, Michael
  Krebs, Miche Baker-Harvey, and Tyler Sanderson.
\newblock Vm live migration at scale.
\newblock {\em ACM SIGPLAN Notices}, 53(3), March 2018.

\bibitem{topdownanalysis.ispass14}
Ahmad Yasin.
\newblock {A Top-Down Method for Performance Analysis and Counters
  Architecture}.
\newblock In {\em IEEE International Symposium on Performance Analysis of
  Systems and Software (ISPASS)}, 2014.

\bibitem{tmam.web21}
{Top-down Microarchitecture Analysis Method}.
\newblock
  \url{https://software.intel.com/content/www/us/en/develop/documentation/vtune-cookbook/top/methodologies/top-down-microarchitecture-analysis-method.html},
  2021.

\bibitem{jarus2016top}
Mateusz Jarus and Ariel Oleksiak.
\newblock {Top-Down Characterization Approximation Based on Performance
  Counters Architecture for AMD Processors}.
\newblock {\em Simulation Modelling Practice and Theory}, 2016.

\bibitem{akiyama2017quantitative}
Soramichi Akiyama and Takahiro Hirofuchi.
\newblock {Quantitative Evaluation of Intel PEBS Overhead for Online
  System-noise Analysis}.
\newblock In {\em Proceedings of the 7th International Workshop on Runtime and
  Operating Systems for Supercomputers (ROSS)}, 2017.

\bibitem{hemem.sosp21}
Amanda Raybuck, Tim Stamler, Wei Zhang, Mattan Erez, and Simon Peter.
\newblock {HeMem: Scalable Tiered Memory Management for Big Data Applications
  and Real NVM}.
\newblock In {\em Proceedings of the 28th ACM Symposium on Operating Systems
  Principles (SOSP)}, 2021.

\bibitem{pedregosa2011scikit}
Fabian Pedregosa, Ga{\"e}l Varoquaux, Alexandre Gramfort, Vincent Michel,
  Bertrand Thirion, Olivier Grisel, Mathieu Blondel, Peter Prettenhofer, Ron
  Weiss, Vincent Dubourg, et~al.
\newblock {Scikit-learn: Machine Learning in Python}.
\newblock {\em Journal of Machine Learning Research (JMLR)}, 12, 2011.

\bibitem{ke2017lightgbm}
Guolin Ke, Qi~Meng, Thomas Finley, Taifeng Wang, Wei Chen, Weidong Ma, Qiwei
  Ye, and Tie-Yan Liu.
\newblock {Lightgbm: A Highly Efficient Gradient Boosting Decision Tree}.
\newblock {\em Advances in Neural Information Processing Systems (NIPS)}, 2017.

\bibitem{onnx}
{ONNX}.
\newblock {Open Neural Network Exchange: the Open Standard for Machine Learning
  Interoperability}.
\newblock \url{https://onnx.ai/}, 2021.

\bibitem{olston2017tensorflow}
Christopher Olston, Noah Fiedel, Kiril Gorovoy, Jeremiah Harmsen, Li~Lao,
  Fangwei Li, Vinu Rajashekhar, Sukriti Ramesh, and Jordan Soyke.
\newblock {Tensorflow-serving: Flexible, High-performance ML Serving}.
\newblock In {\em Proceedings of the 31st Conference on Neural Information
  Processing Systems (NIPS)}, 2017.

\bibitem{crankshaw2017clipper}
Daniel Crankshaw, Xin Wang, Guilio Zhou, Michael~J. Franklin, Joseph~E.
  Gonzalez, and Ion Stoica.
\newblock {Clipper: A Low-Latency Online Prediction Serving System}.
\newblock In {\em Proceedings of the 14th USENIX Symposium on Networked Systems
  Design and Implementation (NSDI)}, 2017.

\bibitem{microsoft_trace_analysis}
Eli Cortez, Anand Bonde, Alexandre Muzio, Mark Russinovich, Marcus Fontoura,
  and Ricardo Bianchini.
\newblock {Resource Central: Understanding and Predicting Workloads for
  Improved Resource Management in Large Cloud Platforms}.
\newblock In {\em Proceedings of the 26th ACM Symposium on Operating Systems
  Principles (SOSP)}, 2017.

\bibitem{burns2016borg}
Brendan Burns, Brian Grant, David Oppenheimer, Eric Brewer, and John Wilkes.
\newblock {Borg, Omega, and Kubernetes}.
\newblock {\em Communications of the ACM}, 59(5), 2016.

\bibitem{delimitrou2013paragon}
Christina Delimitrou and Christos Kozyrakis.
\newblock Paragon: Qos-aware scheduling for heterogeneous datacenters.
\newblock {\em ACM SIGPLAN Notices}, 48(4), 2013.

\bibitem{delimitrou2015tarcil}
Christina Delimitrou, Daniel Sanchez, and Christos Kozyrakis.
\newblock {Tarcil: Reconciling Scheduling Speed and Quality in Large Shared
  Clusters}.
\newblock In {\em Proceedings of the 6th ACM Symposium on Cloud Computing
  (SoCC)}, 2015.

\bibitem{schwarzkopf2013omega}
Malte Schwarzkopf, Andy Konwinski, Michael Abdel-Malek, and John Wilkes.
\newblock {Omega: flexible, scalable schedulers for large compute clusters}.
\newblock In {\em Proceedings of the 2013 EuroSys Conference (EuroSys)}, 2013.

\bibitem{verma2015large}
Abhishek Verma, Luis Pedrosa, Madhukar Korupolu, David Oppenheimer, Eric Tune,
  and John Wilkes.
\newblock {Large-scale Cluster Management at Google with Borg}.
\newblock In {\em Proceedings of the 2015 EuroSys Conference (EuroSys)}, 2015.

\bibitem{redis.web21}
{Redis}.
\newblock \url{https://redis.io}, 2021.

\bibitem{voltdb.web21}
{VoltDB}.
\newblock \url{https://www.voltdb.com}, 2021.

\bibitem{tpch.web21}
{TPC-H Benchmark}.
\newblock \url{http://www.tpc.org/tpch}, 2021.

\bibitem{hibench.web21}
{HiBench: The Bigdata Micro Benchmark Suite}.
\newblock \url{https://github.com/Intel-bigdata/HiBench}, 2021.

\bibitem{gapbs.corr15}
Scott Beamer, Krste Asanovi{\'c}, and David Patterson.
\newblock {The GAP Benchmark Suite}.
\newblock {\em arXiv:1508.03619}, 2015.

\bibitem{parsec3.can16}
Xusheng Zhan, Yungang Bao, Christian Bienia, and Kai Li.
\newblock {PARSEC3.0: A Multicore Benchmark Suite with Network Stacks and
  SPLASH-2X}.
\newblock {\em ACM SIGARCH Computer Architecture News (CAN)}, 44(5), 2017.

\bibitem{speccpu2017.web21}
{SPEC CPU 2017}.
\newblock \url{https://www.spec.org/cpu2017}, 2021.

\bibitem{parsec.pact08}
Christian Bienia, Sanjeev Kumar, Jaswinder~Pal Singh, and Kai Li.
\newblock {The PARSEC Benchmark Suite: Characterization and Architectural
  Implications}.
\newblock In {\em IEEE International Conference on Parallel Architectures and
  Compilation Techniques (PACT)}, 2008.

\bibitem{buckland1994relationship}
Michael Buckland and Fredric Gey.
\newblock {The Relationship Between Recall and Precision}.
\newblock {\em Journal of the American Society for Information Science}, 45(1),
  1994.

\bibitem{awsdisk}
{AWS: optimize disk performance for instance store volumes}.
\newblock
  \url{https://docs.aws.amazon.com/AWSEC2/latest/UserGuide/disk-performance.html}.

\bibitem{pcieats.web09}
{PCI Express Address Translation Services}.
\newblock \url{https://composter.com.ua/documents/ats_r1.1_26Jan09.pdf}, 2009.

\bibitem{clio.web21}
Zhiyuan Guo, Yizhou Shan, Xuhao Luo, Yutong Huang, and Yiying Zhang.
\newblock {Clio: A Hardware-Software Co-Designed Disaggregated Memory System}.
\newblock In {\em Proceedings of the 27th ACM International Conference on
  Architectural Support for Programming Languages and Operating Systems
  (ASPLOS)}, 2022.

\bibitem{zombieland.eurosys18}
Vlad Nitu, Boris Teabe, Alain Tchana, Canturk Isci, and Daniel Hagimont.
\newblock {Welcome to Zombieland: Practical and Energy-efficient Memory
  Disaggregation in a Datacenter}.
\newblock In {\em Proceedings of the 2018 EuroSys Conference (EuroSys)}, 2018.

\bibitem{pberry.hotos19}
Irina Calciu, Ivan Puddu, Aasheesh Kolli, Andreas Nowatzyk, Jayneel Gandhi,
  Onur Mutlu, and Pratap Subrahmanyam.
\newblock {Project PBerry: FPGA Acceleration for Remote Memory}.
\newblock In {\em Proceedings of the 17th Workshop on Hot Topics in Operating
  Systems (HotOS XVII)}, 2019.

\bibitem{dredbox.date16}
K.~Katrinis, D.~Syrivelis, D.~Pnevmatikatos, G.~Zervas, D.~Theodoropoulos,
  I.~Koutsopoulos, K.~Hasharoni, D.~Raho, C.~Pinto, F.~Espina, S.~Lopez-Buedo,
  Q.~Chen, M.~Nemirovsky, D.~Roca, H.~Klos, and T.~Berends.
\newblock {Rack-scale Disaggregated Cloud Data Centers: The dReDBox Project
  Vision}.
\newblock In {\em Design Automation and Test in Europe (DATE)}, 2016.

\bibitem{optmemdisagg}
Jorge Gonzalez, Alexander Gazman, Maarten Hattink, Mauricio G.~Palma, Meisam
  Bahadori, Ruth Rubio-Noriega, Lois Orosa, Madeleine Glick, Onur Mutlu, Keren
  Bergman, and Rodolfo Azevedo.
\newblock {Optically Connected Memory for Disaggregated Data Centers}.
\newblock In {\em IEEE 32nd International Symposium on Computer Architecture
  and High Performance Computing (SBAC-PAD)}, 2020.

\bibitem{opencapi.web21}
{OpenCAPI Consortium}.
\newblock \url{https://opencapi.org/}, 2021.

\bibitem{heteroos.isca17}
Sudarsun Kannan, Ada Gavrilovska, Vishal Gupta, and Karsten Schwan.
\newblock {HeteroOS: OS Design for Heterogeneous Memory Management in
  Datacenters}.
\newblock In {\em Proceedings of the 44th Annual International Symposium on
  Computer Architecture (ISCA)}, 2017.

\bibitem{thermostat.asplos17}
Neha Agarwal and Thomas~F. Wenisch.
\newblock {Thermostat: Application-transparent Page Management for Two-tiered
  Main Memory}.
\newblock In {\em Proceedings of the 22nd ACM International Conference on
  Architectural Support for Programming Languages and Operating Systems
  (ASPLOS)}, 2017.

\bibitem{flatflash.asplos19}
Ahmed Abulila, Vikram~Sharma Mailthody, Zaid Qureshi, Jian Huang, Nam~Sung Kim,
  Jinjun Xiong, and Wen mei Hwu.
\newblock {FlatFlash: Exploiting the Byte-Accessibility of SSDs within a
  Unified Memory-Storage Hierarchy}.
\newblock In {\em Proceedings of the 24th ACM International Conference on
  Architectural Support for Programming Languages and Operating Systems
  (ASPLOS)}, 2019.

\bibitem{autotiering.atc21}
Jonghyeon Kim, Wonkyo Choe, and Jeongseob Ahn.
\newblock {Exploring the Design Space of Page Management for Multi-Tiered
  Memory Systems}.
\newblock In {\em Proceedings of the 2021 USENIX Annual Technical Conference
  (ATC)}, 2021.

\bibitem{zswap.web20}
{Linux Memory Management Documentation - zswap}.
\newblock \url{https://www.kernel.org/doc/html/latest/vm/zswap.html}, 2020.

\bibitem{mldistore.mlsys21}
Giulio Zhou and Martin Maas.
\newblock {Learning on Distributed Traces for Data Center Storage Systems}.
\newblock In {\em Proceedings of the 4th Conference on Machine Learning and
  Systems (MLSys)}, 2021.

\bibitem{llama.asplos20}
Martin Maas, David~G. Andersen, Michael Isard, Mohammad~Mahdi Javanmard,
  Kathryn~S. McKinley, and Colin Raffel.
\newblock {Learning-based Memory Allocation for C++ Server Workloads}.
\newblock In {\em Proceedings of the 25th ACM International Conference on
  Architectural Support for Programming Languages and Operating Systems
  (ASPLOS)}, 2020.

\bibitem{sinan.asplos21}
Yanqi Zhang, Weizhe Hua, Zhuangzhuang Zhou, G.~Edward Suh, and Christina
  Delimitrou.
\newblock {Sinan: ML-Based and QoS-Aware Resource Management for Cloud
  Microservices}.
\newblock In {\em Proceedings of the 26th ACM International Conference on
  Architectural Support for Programming Languages and Operating Systems
  (ASPLOS)}, 2021.

\bibitem{nmprefetch.asplos21}
Zhan Shi, Akanksha Jain, Kevin Swersky, Milad Hashemi, Parthasarathy
  Ranganathan, and Calvin Lin.
\newblock {A Hierarchical Neural Model of Data Prefetching}.
\newblock In {\em Proceedings of the 26th ACM International Conference on
  Architectural Support for Programming Languages and Operating Systems
  (ASPLOS)}, 2021.

\bibitem{dlcache.micro19}
Zhan Shi, Xiangru Huang, Akanksha Jain, and Calvin Lin.
\newblock {Applying Deep Learning to the Cache Replacement Problem}.
\newblock In {\em 52nd Annual IEEE/ACM International Symposium on
  Microarchitecture (MICRO-52)}, 2019.

\bibitem{ccnuma.isca97}
James Laudon and Daniel Lenoski.
\newblock {The SGI Origin: A ccNUMA Highly Scalable Server}.
\newblock In {\em Proceedings of the 24th Annual International Symposium on
  Computer Architecture (ISCA)}, 1997.

\bibitem{asymmemplacement.atc15}
Baptiste Lepers, Vivien Quéma, and Alexandra Fedorova.
\newblock {Thread and Memory Placement on NUMA Systems: Asymmetry Matters}.
\newblock In {\em Proceedings of the 2015 USENIX Annual Technical Conference
  (ATC)}, 2015.

\bibitem{corbet2012autonuma}
Jonathan Corbet.
\newblock Autonuma: the other approach to numa scheduling.
\newblock \url{https://lwn.net/Articles/488709/}, 2012.

\bibitem{rao2010vnuma}
Dulloor~Subramanya Rao and Karsten Schwan.
\newblock {vNUMA-mgr: Managing VM memory on NUMA platforms}.
\newblock In {\em IEEE International Conference on High Performance Computing
  (HiPC)}, 2010.

\bibitem{liu2014optimizing}
Ming Liu and Tao Li.
\newblock {Optimizing Virtual Machine Consolidation Performance on NUMA Server
  Architecture for Cloud Workloads}.
\newblock In {\em Proceedings of the 41th Annual International Symposium on
  Computer Architecture (ISCA)}, 2014.

\end{thebibliography}


}
\endgroup



\end{document}